\documentclass[twocolumn,prd,amsmath,amssymb,amsfonts,showpacs]{revtex4}% Physical Review D
\usepackage{graphicx}
\usepackage{hyperref}

\newcommand{\phm}{\hphantom{-}}
\newcommand{\phn}{\hphantom{0}}

\begin{document}

Phys. Rev. D \textbf{76}, 073007 (2007)

\copyright 2007 The American Physical Society

\url{http://link.aps.org/abstract/PRD/v76/e073007}

\title{$Q^2$-evolution of nucleon-to-resonance transition form factors\\ in a QCD-inspired vector-meson-dominance model}

\author{G.~Vereshkov}
\email{gveresh@gmail.com}
\author{N.~Volchanskiy}
\email{nikolay.volchanskiy@gmail.com}
\affiliation{%
Research Institute of Physics\\
Southern Federal University\\
344090, Rostov-na-Donu, Russia
}%
%\date{\today}

\begin{abstract}
We adopt the vector-meson-dominance approach to investigate $Q^2${\nobreakdash-\hspace{0pt}}evolution of $NR${\nobreakdash-\hspace{0pt}}transition form factors ($N$ denotes nucleon and $R$ an excited resonance) in the first and second resonance regions. The developed model is based upon conventional $\gamma NR${\nobreakdash-\hspace{0pt}}interaction Lagrangians, introducing three form factors for spin{\nobreakdash-\hspace{0pt}}$3/2$ resonances and two form factors for spin{\nobreakdash-\hspace{0pt}}$1/2$ nucleon excitations. Lagrangian form factors are expressed as dispersionlike expansions with four or five poles corresponding to the lowest excitations of the mesons $\rho(770)$ and $\omega(782)$. Correct high{\nobreakdash-\hspace{0pt}}$Q^2$ form factor behavior predicted by perturbative QCD is due to phenomenological logarithmic renormalization of electromagnetic coupling constants and linear superconvergence relations between the parameters of the meson spectrum. The model is found to be in good agreement with all the experimental data on $Q^2${\nobreakdash-\hspace{0pt}}dependence of the transitions $N\Delta(1232)$, $NN(1440)$, $NN(1520)$, $NN(1535)$. We present fit results and model predictions for high-energy experiments proposed by JLab. Besides, we make special emphasis on the transition to perturbative domain of $N\Delta(1232)$ form factors.
\end{abstract}

\pacs{13.40.Gp, 12.40.Vv, 14.20.Dh, 14.20.Gk}

\maketitle

%==============================================================================================
%==============================================================================================
%==============================================================================================

\section{Introduction}

Substantial experimental efforts have been made in recent years to measure $Q^2${\nobreakdash-\hspace{0pt}}dependence of baryon transition form factors via resonant inelastic $eN${\nobreakdash-\hspace{0pt}}scattering. In particular, the data on the transitions $N\Delta(1232)$, $NN(1440)$, $NN(1520)$, $NN(1535)$ were obtained up to squared momentum transfer $Q^2 = 6 \text{ GeV}^2$ in exclusive experiments carried out in such facilities as JLab, MIT-Bates, MAMI and others \cite{un-06, fr-99, st-06, sp-05, el-06, ke-05, ar-99, th-01, ku-03, eg-06, ka-97, ba-02, po-01, jo-02, jo-03, jo-04, me-01, ri-03, bi-03, wa-98, vi-02, bl-01} (see also review \cite{bu-04}). Although there is a total absence of exclusive data above $Q^2 = 6 \text{ GeV}^2$, a large body of inclusive measurements by SLAC is available for the resonance $\Delta(1232)$ (up to almost $10 \text{ GeV}^2$) and $N(1535)$ (up to $21 \text{ GeV}^2$) (see \cite{st-98}, review \cite{st-93} and references therein). Besides, new high{\nobreakdash-\hspace{0pt}}$Q^2$ exclusive measurements are proposed by JLab \cite{st-01}. This study of pion and $\eta$-meson electroproduction is supposed to provide a source of information about $Q^2${\nobreakdash-\hspace{0pt}}behavior of the $\Delta(1232)$ and $N(1535)$ multipole moments up to $Q^2 = 14 \text{ GeV}^2$. Also, there is a discussion motivating possible experiments dealing with resonance excitation in the reaction $\pi N \to e^+ e^- N$ \cite{lu-03} that is a crossed channel of pion electroproduction. All these experimental prospects as well as recent numerous high-precision measurements call for careful theoretical examination and interpretation of the data available. Our work is an example of this effort.

A starting point of such an interpretation is the partial wave analysis (PWA) of both non-polarized and polarized exclusive data, which provide $Q^2${\nobreakdash-\hspace{0pt}}dependence of three helicity amplitudes $A_{1/2}(Q^2)$, $A_{3/2}(Q^2)$, $S_{1/2}(Q^2)$ for spin{\nobreakdash-\hspace{0pt}}$3/2$ resonances and two amplitudes $A_{1/2}(Q^2)$, $S_{1/2}(Q^2)$ for spin{\nobreakdash-\hspace{0pt}}$1/2$ baryons (PWA is not possible for inclusive processes. Only the transverse amplitude $A_T = (A_{3/2}^2+A_{1/2}^2)^{1/2}$ can be extracted from inclusive data.) Detailed discussions of amplitude-extraction techniques and their model dependence is given in the reviews \cite{bu-04,pa-07}.

Helicity amplitudes encode information about the space structure of nucleon excitations and physical nature of baryon transition form factors. The underlying fundamental theory of $NR${\nobreakdash-\hspace{0pt}}transitions is quantum chromodynamics (QCD). In the non-perturbative domain, however, \textit{ab initio} QCD calculations are not currently feasible, because of their extreme complexity. Even numeric calculations \cite{LQCD1,LQCD2,LQCD3} utilizing lattice-QCD techniques are hindered by computer power available now. The best presentday lattice calculations are obtained for pion mass above $0.3 \text{ GeV}$ and, consequently, disagree with accelerator measurements. To reconcile them with experiment, one requires additional phenomenological suppositions about extrapolation to the physical masses \cite{pa-07,th-03}. In the perturbative domain, QCD (pQCD) calculations provide important and reliable predictions about asymptotic (high{\nobreakdash-\hspace{0pt}}$Q^2$) behavior of resonance form factors \cite{st-93}, still there is not clear indication of the onset of hard scattering processes in the existing experimental data base (except for the inclusive data on the $NN(1535)${\nobreakdash-\hspace{0pt}}transition \cite{st-93}).

Since QCD itself is currently not able to provide a comprehensive treatment of resonant helicity amplitudes, a lot of phenomenological models have emerged. Some of the most prominent approaches are: quark shell models such as chiral bag model \cite{chiBM} and numerous constituent quark models (single quark transition model \cite{SQTM}, hypercentral model \cite{hCQM}, model with two-body exchange currents \cite{Buchmann} and so forth); soliton models (Skyrmion models \cite{SkyrmeModel1,SkyrmeModel2}, linear $\sigma$-model \cite{sigmaModel,CDMsigmaModel1,CDMsigmaModel2}, chiral chromodielectric model \cite{CDMsigmaModel1,CDMsigmaModel2}, chiral quark-soliton model \cite{chiQSM}); algebraic approach \cite{slaughter-04}; generalized parton distributions (see reviews \cite{GPD1,GPD2,GPD3,GPD4,pa-07}); chiral effective field theories \cite{ga-06,chiEFT2}. The majority of these models are in quantitative agreement with the experimental data points just at small $Q^2 < 1.5 \text{ GeV}^2$.

In this paper our main objective is to demonstrate that effective-field theory incorporating vector-meson-dominance (VMD) effects can reproduce $Q^2${\nobreakdash-\hspace{0pt}}evolution of resonant helicity amplitudes in both perturbative and non-perturbative domains. So far VMD models have been successfully applied to address the same problem of $Q^2${\nobreakdash-\hspace{0pt}}evolution mostly in elastic $eN${\nobreakdash-\hspace{0pt}}scattering \cite{gk-85,gk-92,ia-04a,ia-04b,ia-04c,lo-01,lo-02}, the lowest-resonance electroproduction \cite{WanIachello,Devenish1976}, and deeply virtual Compton scattering \cite{Capua2007}. All the versions of the VMD model contain several ambiguities:
\begin{enumerate}
\item \textit{The choice of the vector-meson spectrum.} The majority of the models take into account only the lightest vector mesons $\rho(770)$, $\omega(782)$, $\phi(1020)$ \cite{gk-85,gk-92,ia-04a,ia-04b,ia-04c,du-02} and seldom $\rho(1450)$ \cite{lo-01,lo-02}, $\omega(1420)$ \cite{lo-02}. Only the model \cite{du-02} includes all the vector mesons reported by experimentalists before its publication.
\item \textit{The choice of the interaction Lagrangian.} The problem of the interaction-Lagrangian  symmetries becomes important in the description of high-spin resonances. First of all, most Lagrangians currently in use could break free-field subsidiary constraints reducing nonphysical degrees of freedom of the Rarita-Schwinger field \cite{ra-sh}, which result in different pathologies, e.g., excitation of superluminal modes \cite{ve-zw}. This could be avoided in the theory with additional symmetries such as gauge invariance of the resonance field \cite{pa-ti}.
\item \textit{The way to impose high-$Q^2$ behavior predicted by pQCD on Lagrangian form factors.} In the models involving only ground-state vector mesons and their first excitations, agreement with pQCD is due to artificial suppression of meson-spectrum parameters by power corrections \cite{gk-85,gk-92,ia-04a,ia-04b,ia-04c,lo-01,lo-02}. Obviously, such a suppression of photon-meson couplings disagrees with quantum field theory, since parameters of vector-meson spectrum can be renormalized by only slight logarithmic functions. Another way to fulfill asymptotic pQCD-constraints is by linear superconvergence relations between meson parameters \cite{du-03}.
\item \textit{The way to treat logarithmic corrections.} Logarithmic dependencies in form factors are an essential feature of the model. They are necessary to incorporate logarithmic corrections to pQCD-asymptotes, though not calculated directly yet.
\item \textit{Inclusion of continuum contributions.} An infinite number of virtual intermediate multihadron states give rise to continuum contributions to form factors. Most VMD models include only the $2\pi$-cut associated with the lightest isovector intermediary state \cite{ia-04a,ia-04b,ia-04c}. Also some models take into account $K \bar K$ and $\rho \pi$ continua \cite{be-06} or $3\pi$-continuum and effective inelastic cuts \cite{du-02}.
\end{enumerate}

In this paper we build up a VMD model satisfying asymptotic constraints predicted by pQCD (that's why, we refer to it as ``QCD-inspired''). In the perturbative domain, QCD expects resonant helicity amplitudes to have power-logarithmic asymptotes and fall faster than the dipole (ground-state-meson) model predicts. We prefer to impose correct asymptotic behavior on form factors by superconvergence relations in the manner of the paper \cite{du-02}, rather than by invoking unphysical power suppression. As we show in Sec. \ref{MVDM section}, this requires the model to include at least four vector mesons. Nevertheless, this does not lead to a dramatic increase in the number of free parameters. For example, in the simplest four-meson model, vector-meson spectrum comprises only one independent parameter, and even this model is in accord with the data at all $Q^2$ with high accuracy (except for the resonance $N(1440)$, whose peculiar structure can be reproduced in the model with at least five mesons). Our VMD model differs, however, from that of Ref.~\cite{du-02}, because the significant feature of our calculation is phenomenological logarithmic renormalization of the parameters of the vector-meson spectrum. Logarithmic renormalization is essential to comply with both power and logarithmic pQCD-behavior, which we discuss in Sec. \ref{QHD}. Another difference of this work from Ref.~\cite{du-02} is that we neglect continuum contributions to transition form factors, since this simplest (tree-level) parametrization is found to describe satisfactorily all the experimental data. The following discussion is constrained to the calculation of the vector transition form factors for the first four low-lying baryon resonances. Application of the model to description of the nucleon axial and elastic form factors is the topic of our further publications~\cite{ve-07}.

In this paper we make use of the traditional Lagrangian (Eqs.~\eqref{M32}, \eqref{G32}) of the $\gamma^*NR${\nobreakdash-\hspace{0pt}}interaction for spin{\nobreakdash-\hspace{0pt}}$3/2$ resonances (see, e.g., \cite{da-91,fe-mo}). This coupling excites low-spin background of the Rarita-Schwinger spin-vector field. But, despite this mathematical inconsistency, it is intensively utilized in helicity-amplitude extraction \cite{fr-99,fe-mo} as well as theoretical studies of $Q^2${\nobreakdash-\hspace{0pt}}evolution of form factors \cite{la-06,ga-06}. In the following, we confine ourselves to working with only this inconsistent but popular interaction, since it enables us to demonstrate the validity of VMD approach in physics of transition form factors at the entire range of $Q^2$. We are going, however, to discuss alternative couplings in our further publication \cite{futurepub}.

The remainder of this paper proceeds as follows. Section~\ref{Lagrangians and asymptotes} comprises Lagrangians and corresponding to them cross-section formulas. Also we present a detailed discussion of how to bring an effective-field-theory model into accordance with pQCD-predictions. The next Sec.~\ref{MVDM section} lays out our VMD model, including superconvergence relations and logarithmic renormalization. Section \ref{fitsection} contains fits, model predictions, and discussion of these results with the emphasis on the transition to pQCD regime of the $N\Delta(1232)$ form factors. Finally, Sec.~\ref{Conclusion} is a summary of our main conclusions as well as possible areas of extension of the model and improvement to it. The technical details of our calculations, concerning the choice of helicity-amplitude signs, can be found in the appendix.

%==============================================================================================
%==============================================================================================
%==============================================================================================

\section{Phenomenological model of baryon electroproduction}\label{Lagrangians and asymptotes}

In this rather long section we write down conventional $\gamma NR$-vertexes and define resonant helicity amplitudes by their relations to observables, i.e., differential cross sections. Starting from these formulas, it is quite easy to compute relations between phenomenological-model form factors, comprised in the $\gamma NR$-vertexes \eqref{G32} and \eqref{G12}, and helicity amplitudes (or any other quantities traditionally used to describe resonant $eN$-scattering). However, an important step in the calculations is the choice of amplitude phases, which is discussed in the appendix. It should be stressed that the choice of amplitude phases could strongly influence the quality of fits to experimental data. Besides, to determine amplitude phases is not a straightforward task if one uses simple factorized cross-section formulas, not involving amplitude interference, as it is, e.g., in Ref.~\cite{la-06} and this paper (see the appendix).

Also in this section we extract Lagrangian form factors from the experimental data on helicity amplitudes and discuss their asymptotic behavior.

%==============================================================================================
%==============================================================================================
%==============================================================================================

\subsection{Matrix elements and vertex operators}

To discuss the underlying physics of the nucleon-to-resonance transition form factors, it seems reasonable to combine the results and approaches of quantum chromodynamics and effective-field-theory (EFT) models. Moreover, this synthesis is mathematically inescapable. As quark confinement is an essential feature of QCD, any amplitude of physical process presented by a functional integral over the space of the quark and the gluon fields can be equally expressed as the integral over hadron degrees of freedom \cite{vo-06}. This problem, however, is too complicated to be applied directly in the nonperturbative domain of QCD. In such a situation it is EFT that exposes limitations of the vertexes of effective hadron interactions~--- the objects of QCD calculations. For example, EFT provides the matrix elements for the electroproduction of spin{\nobreakdash-\hspace{0pt}}$3/2$ baryon resonances:
\begin{align}\label{M32}
	M(\gamma^*N\to R)=&\langle R |\bar
	u_R^\mu(p')\Gamma_{\mu\nu\lambda}(q,p,p') \times \nonumber\\
	&\times \left(q^\nu e^\lambda(q)-q^\lambda e^\nu(q)\right)u_N(p)|N\rangle,
\end{align}
where $q = p'-p$ is the 4-momentum transfer; $u_R^\mu(p'),\,u_N(p)$ are the resonance vector-spinor and the nucleon spinor; $e^\nu(q)$ is the photon polarization; $\Gamma_{\mu\nu\lambda}(q,p,p')$ is the vertex operator, which is antisymmetric on the last two indices. To write the matrix element for the transition $N \to$ spin-$1/2$ resonance, one should just omit the Lorentz index $\mu$ in Eq. (\ref{M32}):
\begin{align}\label{M12}
	M(\gamma^*N\to R)=&\langle R |\bar
	u_R(p')\Gamma_{\nu\lambda}(q,p,p') \times \nonumber\\
	&\times \left(q^\nu e^\lambda(q)-q^\lambda e^\nu(q)\right)u_N(p)|N\rangle.
\end{align}

The next important step to build up an EFT model of baryon electroproduction is to decompose vertexes $\Gamma_{\mu\nu\lambda}(q,p,p')$, $\Gamma_{\nu\lambda}(q,p,p')$ in terms of the particular spin-tensor basis. The scalar coefficients of this expansion are form factors, their number being equal to the number of basic elements. We note that the basis should be postulated in both QCD and EFT. In QCD the decomposition made in terms of quark correlators yields directly form factors in the domain of perturbative QCD \cite{br-06}. In EFT form factors can be evaluated by means of either the dispersion relation approach or VMD model. The linking idea of these two methods~--- QCD and EFT~--- is quark-hadron duality, i.e., the form factor asymptotes calculated in both pQCD and EFT must be the same.

To understand the physical origin of electromagnetic form factors, one should address two aspects of the problem. First of all, the reliable physical arguments fixing mathematical structure of the vertexes $\Gamma_{\mu\nu\lambda}(q,p,p')$, $\Gamma_{\nu\lambda}(q,p,p')$ should be discussed. In the phenomenology of spin-vector resonance electroproduction the following vertex is often in use \cite{la-06,ga-06}:
\begin{align}\label{G32}
	\Gamma_{\mu\nu\lambda}&(q,p,p') = \frac12 (g_{\mu\nu}g_{\lambda\sigma}-g_{\mu\lambda}g_{\nu\sigma})
	\times {} \nonumber \\ &{} \times
	\left( \frac{C_3(q^2)}{M_N}\gamma^\sigma + \frac{C_4(q^2)}{M_N^2}p^{\sigma '}+\frac{C_5(q^2)}{M_N^2}p^{\sigma}\right)\gamma_R,
\end{align}
where $C_3(q^2),\, C_4(q^2),\, C_5(q^2)$ are phenomenological form factors in the notation of Ref.~\cite{la-06}; $\gamma_R = \gamma_5$ for $R=\Delta(1232)$ and $\gamma_R =1$ for $R=N^*(1520)$. From this point on, we label form factors in such a manner to unify the notation of the non-spin-flip and spin-flip amplitudes:
\begin{gather*}
	C_3(Q^2) \equiv F_1(Q^2),\qquad C_4(Q^2) \equiv F_2(Q^2),\\ C_5(Q^2) \equiv F_3(Q^2)
\end{gather*}
(this notation is similar to that of the elastic Dirac $F_1(Q^2)$ and Pauli $F_2(Q^2)$ form factors).

The theory of the $J=1/2$ resonance electroproduction is based upon the vertex \cite{az-07, la-06}
\begin{align}\label{G12}
	\Gamma_{\nu\lambda}&(q) = \nonumber\\
	&\displaystyle\frac12\left(\frac{G_1(q^2)}{M^2}(\gamma_\nu q_\lambda-\gamma_\lambda q_\nu)-
	\frac{G_2(q^2)}{M} \sigma_{\nu\lambda}\right)\gamma_R,
\end{align}
where $\sigma_{\nu\lambda} = \frac12 (\gamma_\nu \gamma_\lambda - \gamma_\lambda \gamma_\nu)$; $\gamma_R = \gamma_5$ for $R=N(1535)$ and $\gamma_R =1$ for $R=N(1440)$; $G_1(Q^2)$, $G_2(Q^2)$ are, respectively, the non-spin-flip and spin-flip form factors; $M=M_N$ is the normalization factor (there is another convention $M=M_R+M_N$ \cite{la-06} but, in our opinion, it is inconvenient especially when the baryons are off the mass shell).

It should be noted that there are two kinds of $Q^2$-dependent functions in any definition of the vertex operator. The functions of the first kind are multiplicative factors fixed by the structure of the vertex itself. The second kind is form factors $F_\alpha(Q^2),\, G_\alpha(Q^2)$. If one treats form factors as just phenomenological objects, Eqs. (\ref{G32}) and (\ref{G12}) define the general model since three arbitrary functions $F_1(Q^2)$, $F_2(Q^2)$, $F_3(Q^2)$ are used to describe three observables $A_{1/2}(Q^2)$, $A_{3/2}(Q^2)$, $S_{1/2}(Q^2)$ and two functions $G_1(Q^2)$, $G_2(Q^2)$ are used to describe two amplitudes $A_{1/2}(Q^2)$, $S_{1/2}(Q^2)$. But to evaluate form factors in the framework of any particular dynamics model, care must be taken in choosing the first kind functions dependent on kinematic variables. However, in this paper we deal with only conventional models defined by Eqs. (\ref{G32}) and (\ref{G12}) and skip the discussion of their mathematical structure.

The second aspect of the physical origin of form factors is the modeling the functions $F_\alpha(Q^2),\, G_\alpha(Q^2)$. In this paper, to evaluate form factors, we adopt the VMD approach. The agreement of this model with quark-hadron duality (i.e., pQCD asymptotic behavior) is due to the superconvergent relations between meson-spectrum parameters and logarithmic renormalization of effective coupling parameters.

%==============================================================================================
%==============================================================================================
%==============================================================================================

\subsection{Helicity amplitudes and cross sections --- notation}

The pairs of Eqs. (\ref{M32}), (\ref{G32}) and (\ref{M12}), (\ref{G12}) allow to compute photoabsorption amplitudes. The differential cross section of the on-shell resonance electroproduction is expressed in terms of these amplitudes as follows:
\begin{align}\label{sig}
	\frac{d\sigma}{dQ^2} (eN &\to eR) =
	\frac{\alpha M_N(M_R^2-M_N^2)}{2Q^2(s-M_N^2)^2 (1-\varepsilon)} \times {} \nonumber\\
	& {} \times \left[ 2|\tilde S_{1/2}(Q^2)|^2 \varepsilon(s,Q^2) + |A_{T}(Q^2)|^2 \right],
\end{align}
where
\begin{align*}
	\varepsilon(s,&Q^2) = \\
	& \left\{1+\frac{Q^4+2Q^2(M_R^2+M_N^2)+(M_R^2-M_N^2)^2}
	{2\left[(s-M_N^2)(s-M_R^2)-sQ^2\right]}\right\}^{-1}
\end{align*}
is the virtual photon polarization parameter;
\begin{align*}
	|\tilde S_{1/2}(Q^2)| = |S_{1/2}(Q^2)|\left[1+\frac{(Q^2+M_R^2-M_N^2)^2}{4M_N^2Q^2}\right]^{-1/2}
\end{align*}
is the amplitude for the absorption of a longitudinally polarized photon in the normalization that we use from now on. The transverse helicity amplitude is $A_T(Q^2) \equiv A_{1/2}(Q^2)$ for spin-$1/2$ resonances and
\begin{equation} \label{AT}
	|A_T(Q^2)|^2=|A_{1/2}(Q^2)|^2+|A_{3/2}(Q^2)|^2
\end{equation}
for spin-$3/2$ resonances. The squared magnitudes of the helicity amplitudes and the cross sections for the absorption
of transverse and longitudinal photons are equal up to a numeric factor.

The simplest but approximate cross section of the off-shell electroproduction is
\begin{widetext}\label{sigW}
\begin{multline}
	\frac{d^2\sigma}{dQ^2dW^2}(eN\to eR\to eN+\text{mesons}) =
	\frac{\alpha M_N (W^2-M_N^2)}{2Q^2(s-M_N^2)^2[1-\varepsilon(s,Q^2,W^2)]}
	\times {} \\ \times
	\left\{ 2|\tilde S_{1/2}(Q^2,W^2)|^2\varepsilon(s,Q^2,W^2)+|A_{T}(Q^2,W^2)|^2 \right\}
	\frac{\pi^{-1}M_R \Gamma_R}{(W^2-M_R^2)^2+M_R^2\Gamma_R^2},
\end{multline}
\end{widetext}
where
\[
\displaystyle W^2=\left(\sum_ap'_{(a)}\right)^2
\]
is the squared invariant mass of the final hadron state; $M_R$, $\Gamma_R$ are the Breit-Wigner mass and the total width of resonance;
\begin{align}
	\varepsilon &(s,Q^2,W^2) = {} \nonumber \\
	&\left[ 1+\frac{Q^4+2Q^2(W^2+M_N^2)+(W^2-M_N^2)^2}
	{2\left[(s-M_N^2)(s-W^2)-sQ^2\right]} \right]^{-1}.
\end{align}

In addition to helicity amplitudes for $\Delta(1232)$, we will also use magnetic dipole form factor in the Jones-Scadron convention \cite{jo-sc}
\begin{align*}
	G_\text{M}^* &(Q^2) =
	\\
	&- \left[ \frac{M_N^3 (M_\Delta^2-M_N^2)}{2 \pi \alpha (M_\Delta+M_N)^2} \right]^{1/2} \frac{A_{1/2} + \sqrt{3} A_{3/2}}{\left[ Q^2+(M_\Delta-M_N)^2 \right]^{1/2}},
\end{align*}
the ratio $R_\text{EM}$ between electric quadrupole and magnetic dipole multipoles
\begin{align*}
	R_\text{EM} (Q^2) = - \frac{G_\text{E}^*}{G_\text{M}^*} = \frac{A_{1/2}-\displaystyle \frac{1}{\sqrt{3}} A_{3/2}}{A_{1/2}+ \sqrt{3} A_{3/2}},
\end{align*}
and the ratio $R_\text{SM}$ of Coulomb quadrupole multipole to magnetic dipole one
\begin{align*}
	R_\text{SM} (Q^2) = - \frac{\vert \mathbf{q} \vert}{2 M_\Delta} \frac{G_\text{C}^*}{G_\text{M}^*} = \frac{\sqrt{2} S_{1/2}}{A_{1/2}+ \sqrt{3} A_{3/2}},
\end{align*}
where $\mathbf{q}$ is the photon 3-momentum with the modulus $\vert \mathbf{q} \vert = Q^+ Q^- / 4 M$, $Q^\pm = \left[ Q^2 + \left( M_\Delta \pm M_N \right)^2 \right]^{1/2}$, $M=M_\Delta$ in the rest frame of the $\Delta$ and $M=M_N$ in the laboratory frame, in which the initial nucleon is at rest. Note that the amplitude $S_{1/2}$ is not a Lorentz scalar in the convention utilized by experimentalists \cite{gorchtein:055202}. In the following we use the lab frame to calculate this quantity.

%==============================================================================================
%==============================================================================================
%==============================================================================================

\subsection{Helicity amplitudes and extracted form factors}

%==============================================================================================
%==============================================================================================
%==============================================================================================

\subsubsection{The $P_{33}(1232),\; D_{13}(1520) $}

The amplitudes for the electroproduction of spin-$3/2$ resonances calculated within the model (\ref{G32}) are
\begin{widetext}
\begin{align}\label{A32}
	 A_{3/2}(Q^2)
	 = 
	 {}&{}\mp \left[ \frac{\pi\alpha(Q^2+(M_R\mp M_N)^2)}{M_N^5(M_R^2-M_N^2)}\right]^{1/2}
	 \biggl[ M_N \left(M_R\pm M_N\right) F_1(Q^2) + {}
	 \nonumber \\
	 &+ \frac12\left(M_R^2-M_N^2-Q^2\right)F_2(Q^2)+ \frac12\left(M_R^2-M_N^2+Q^2\right)F_3(Q^2) \biggr],
\end{align}
\begin{align}\label{A12}
	 A_{1/2}(Q^2) =
	 {}&{}-\frac{1}{\sqrt{3}} \left[ \frac{\pi\alpha(Q^2+(M_R \mp M_N)^2)}{M_N^5(M_R^2-M_N^2)} \right]^{1/2}
	 \biggl[ \frac{M_N}{M_R}(Q^2\pm M_N(M_R\pm M_N))F_1(Q^2)-
	 \nonumber\\
	 &-\frac12\left(M_R^2-M_N^2-Q^2\right)F_2(Q^2)
	 -\frac12\left(M_R^2-M_N^2+Q^2\right)F_3(Q^2) \biggr],
\end{align}
\begin{align}\label{S12}
	\tilde S_{1/2}(Q^2) & {} \equiv S_{1/2}(Q^2) \left[1+\frac{(Q^2+M_R^2-M_N^2)^2}{4M_N^2Q^2} \right]^{-1/2}
	= \pm \sqrt{\frac23}\left[\frac{\pi\alpha(Q^2+(M_R\mp M_N)^2)} {M_N^5(M_R^2-M_N^2)}\right]^{1/2} \times
	\nonumber \\ &{} \times
	Q \left[M_N F_1(Q^2)+M_R F_2(Q^2)+ \frac{Q^2+M_R^2+M_N^2}{2M_R}F_3(Q^2)\right].
\end{align}
In Eqs. (\ref{A32})--(\ref{S12}) the top signs in $\pm$ and $\mp$ refer to the case of the $\Delta(1232)$, while the bottom ones are for the $N(1520)$. The phases of the amplitudes are chosen under some extra assumptions (see the appendix). The amplitudes $A_{3/2}(Q^2,W^2)$, $A_{1/2}(Q^2,W^2)$, $\tilde S_{1/2}(Q^2,W^2)$ for the off-shell electroproduction can be obtained from Eqs. (\ref{A32})--(\ref{S12}) by the substitution $M_R \to W$. To extract the form factors from the experimental data on photoabsorption amplitudes, one should simply resolve Eqs. (\ref{A32})--(\ref{S12}) with respect to $F_1(Q^2)$, $F_2(Q^2)$, $F_3(Q^2)$. The result is
\begin{align}\label{F1f}
	F_1(Q^2) =
	-\left[\frac{M_N^5(M_R^2-M_N^2)}{\pi\alpha[Q^2+(M_R\mp M_N)^2]}\right]^{1/2}
	\frac{M_R[\pm A_{3/2}(Q^2)+\sqrt{3} A_{1/2}(Q^2)]}{M_N[Q^2+(M_R\pm M_N)^2]},
\end{align}
\begin{align}\label{F2f}
	F_2(Q^2)=
	\pm&\left[\frac{M_N^5(M_R^2-M_N^2)}{\pi\alpha[Q^2+(M_R\mp M_N)^2]}\right]^{1/2}
	\frac{2}{[Q^2+(M_R+M_N)^2][Q^2+(M_R-M_N)^2]}\times{}
	\nonumber \\
	\times \Biggl[ &\bigl[ Q^2+(M_R\mp M_N)^2 \bigr] A_{3/2}(Q^2) +
	M_R M_N \bigl[\pm A_{3/2}(Q^2)-\sqrt{3} A_{1/2}(Q^2)\bigr]+
	\nonumber \\
	&+ \sqrt{\frac32}\frac{M_R\tilde S_{1/2}(Q^2)}{Q} \bigl(Q^2+M_R^2-M_N^2\bigr) \Biggr],
\end{align}
\begin{align}\label{F3f}
	F_3(Q^2) = {} &
	\left[\frac{M_N^5(M_R^2-M_N^2)}{\pi\alpha[Q^2+(M_R\mp M_N)^2]}\right]^{1/2}
	\frac{2M_R^2}{[Q^2+(M_R+M_N)^2][Q^2+(M_R-M_N)^2]}\times
	\nonumber \\
	& {} \times \left[\sqrt{3} A_{1/2}(Q^2)\mp A_{3/2}(Q^2)\pm\sqrt{\frac32}
	\frac{\tilde S_{1/2}(Q^2)}{M_R Q}(Q^2-M_R^2+M_N^2)\right].
\end{align}
\end{widetext}

\begin{figure*}
	\includegraphics[width=0.49\linewidth]{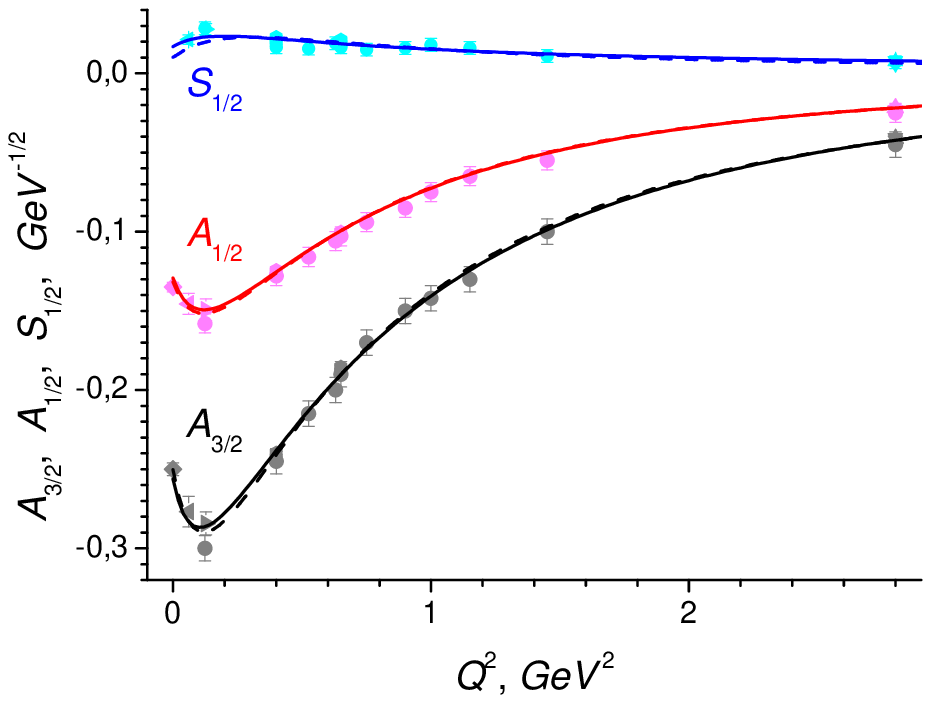}
	\includegraphics[width=0.49\linewidth]{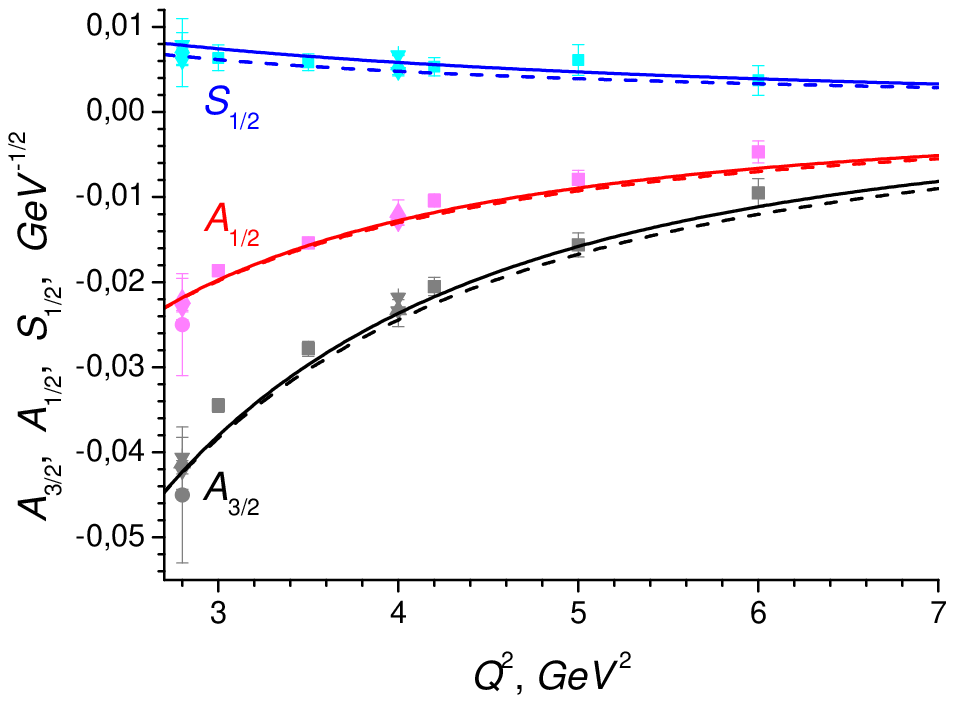}
	\includegraphics[width=0.49\linewidth]{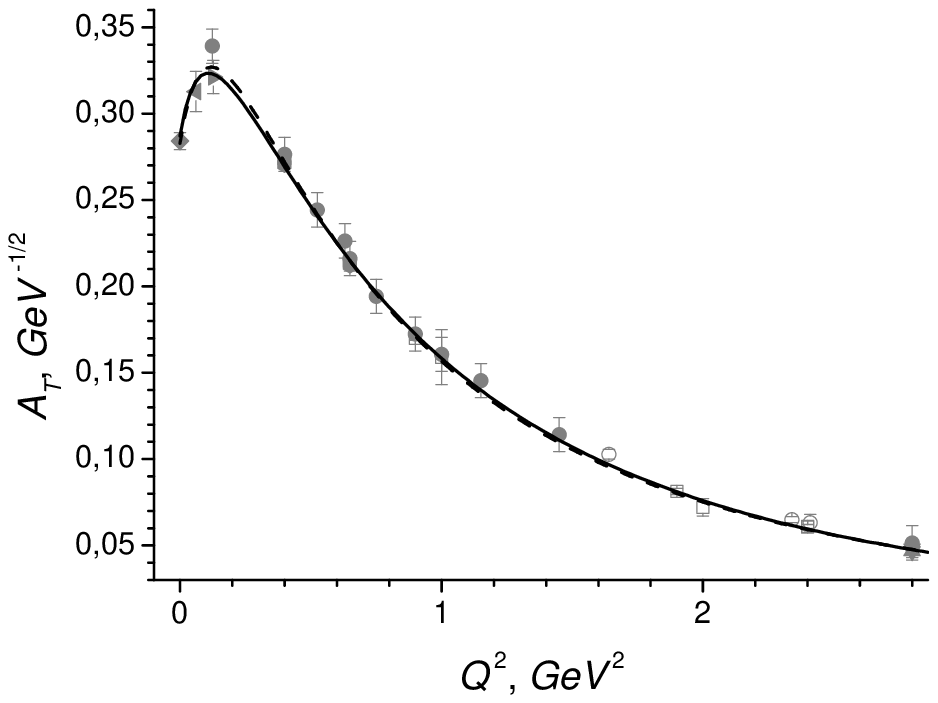}
	\includegraphics[width=0.49\linewidth]{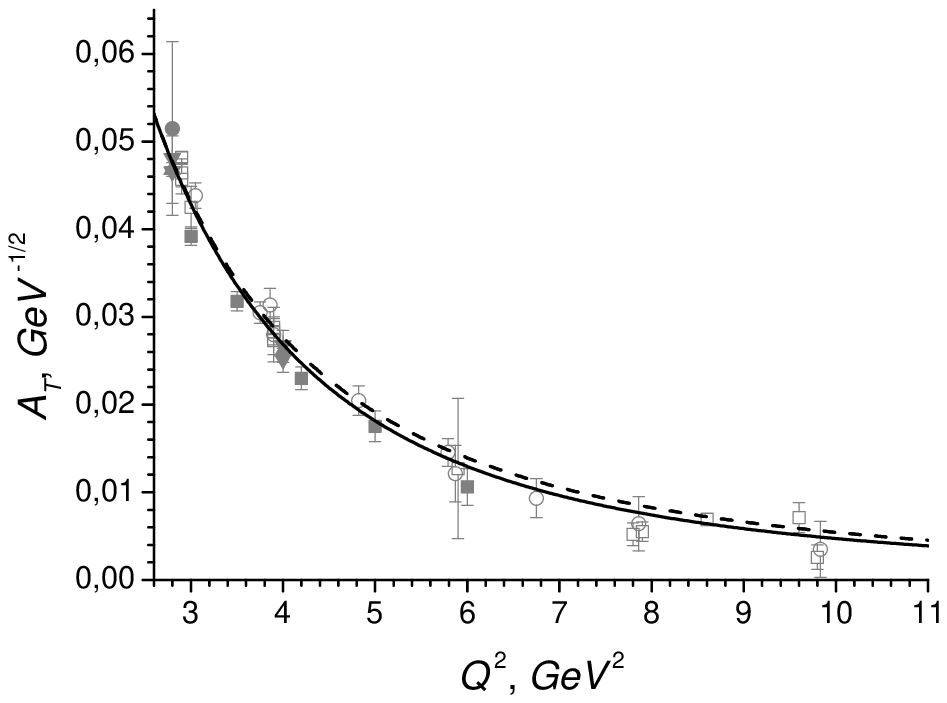}
	\caption{\label{fig:Delta(1232)AAS}Helicity amplitudes of the transition $\gamma^*N \to \Delta(1232)$. The dashed curves correspond to fit F1 with one-parameter logarithmic renormalization \eqref{Li1}, the solid curves to fit F2 with two-parameter renormalization \eqref{Li2}. The data points are denoted as follows: $\blacklozenge$ \cite{PDG}, {\large \textbullet} \cite{ti-04}, {\large $\circ$} \cite{st-98}, {\footnotesize $\blacksquare$} \cite{un-06}, {\footnotesize $\square$} \cite{st-93}, $\blacktriangleleft$ \cite{st-06}, $\blacktriangleright$ \cite{sp-05}, $\blacktriangle$ \cite{fr-99}, $\blacktriangledown$ \cite{ka-01}, $\bigstar$ \cite{ke-05}, filled pentagon \cite{el-06}, filled hexagon \cite{az-05a}.}
\end{figure*}

\begin{figure*}
	\includegraphics[width=0.49\linewidth]{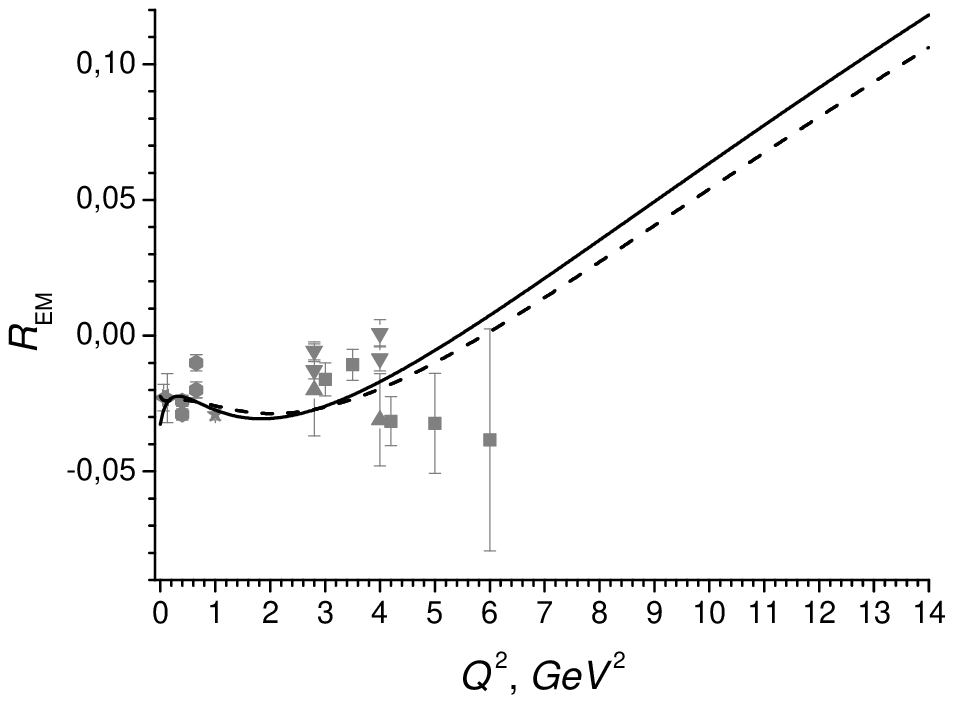}
	\includegraphics[width=0.49\linewidth]{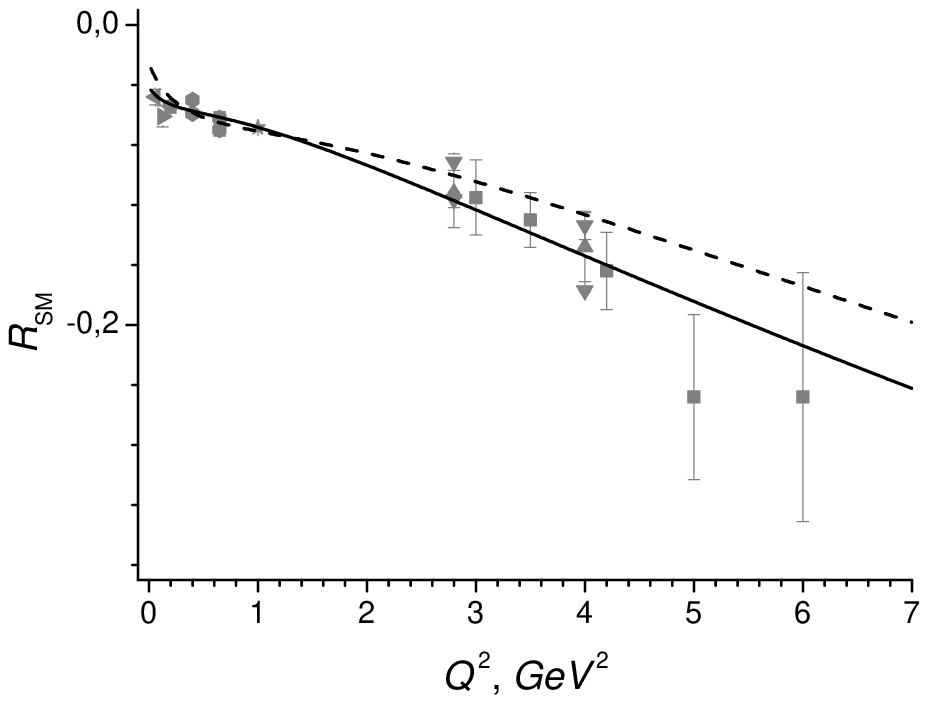}
	\caption{\label{fig:Delta(1232)REMRSM}Ratios $R_\text{EM}$ and $R_\text{SM}$ for the $\Delta(1232)$. Fit curves for $R_\text{EM}$ are prolonged in the domain of proposed upgraded-JLab experiments. Curves and data are the same as in Fig.~\ref{fig:Delta(1232)AAS}.}
\end{figure*}

\begin{figure*}
	\includegraphics[width=8.6cm]{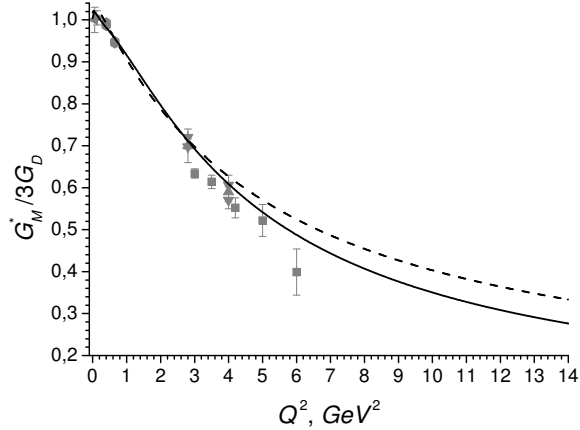}
	\caption{\label{fig:Delta(1232)GM}$\Delta(1232)$ magnetic form factor $G_M^*$ normalized by three times the dipole form factor. We use the standard definition $G_D = (1+Q^2/0.71)^{-2}$. Fit curves are prolonged in the domain of proposed upgraded-JLab experiments. Curves and data are the same as in Fig.~\ref{fig:Delta(1232)AAS}.}
\end{figure*}

\begin{figure*}
	\includegraphics[width=0.49\linewidth]{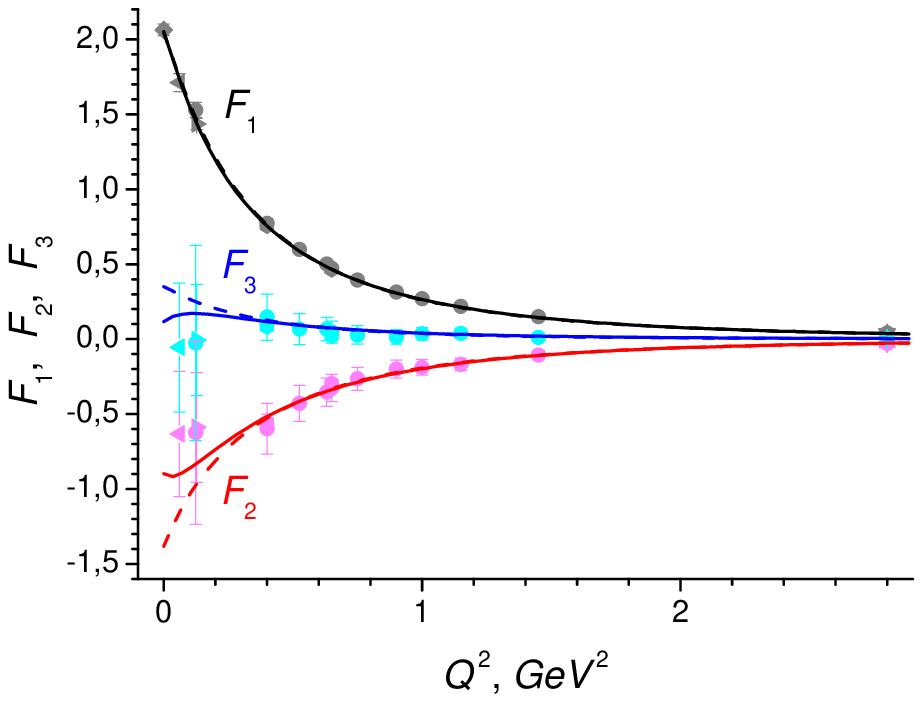}
	\includegraphics[width=0.49\linewidth]{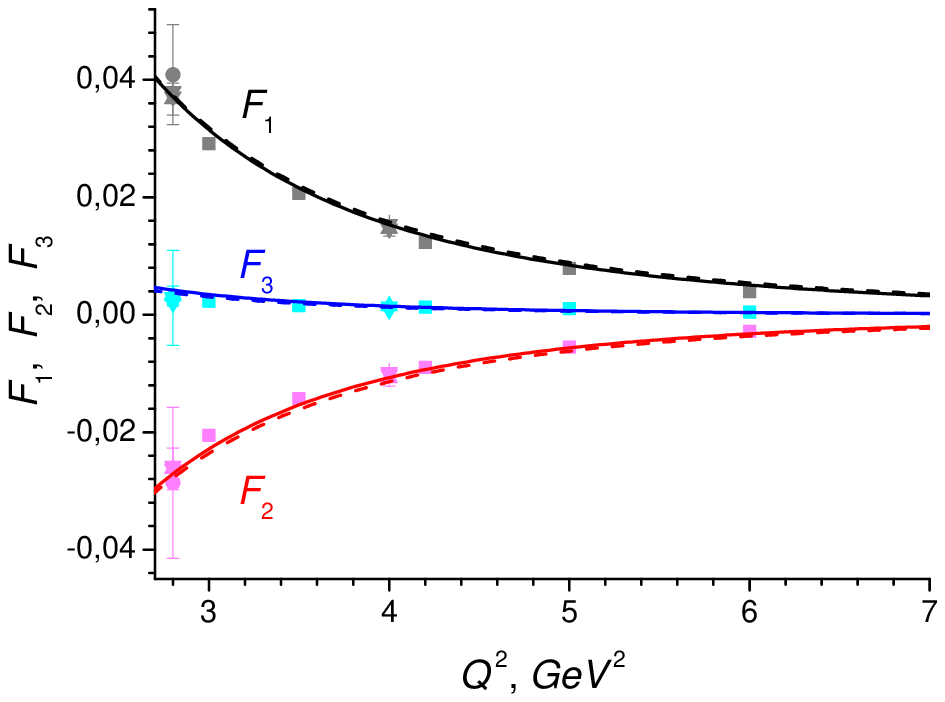}
	\caption{\label{fig:Delta(1232)FFF}Form factors of the model (\ref{G32}) extracted from the data on the transition $\gamma^*N \to \Delta(1232)$. Curves and data are the same as in Fig.~\ref{fig:Delta(1232)AAS}.}
\end{figure*}

\begin{figure*}
	\includegraphics[width=8.6cm]{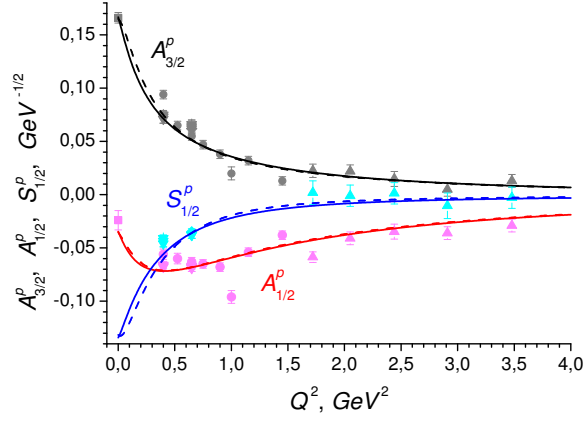}
	\caption{\label{fig:N(1520)AAS} Helicity amplitudes of the transition $\gamma^*p \to N(1520)$. The dashed curves correspond to fit F1 with one-parameter logarithmic renormalization \eqref{Li1}, the solid curves to fit F2 with two-parameter renormalization \eqref{Li2}. The data points are denoted as follows: $\blacklozenge$ \cite{az-05b}, {\large \textbullet} \cite{ti-04}, {\footnotesize $\blacksquare$} \cite{PDG}, $\blacktriangle$ \cite{la-06}, $\blacktriangledown$ \cite{az-05a}.}
\end{figure*}

\begin{figure*}
	\includegraphics[width=0.49\linewidth]{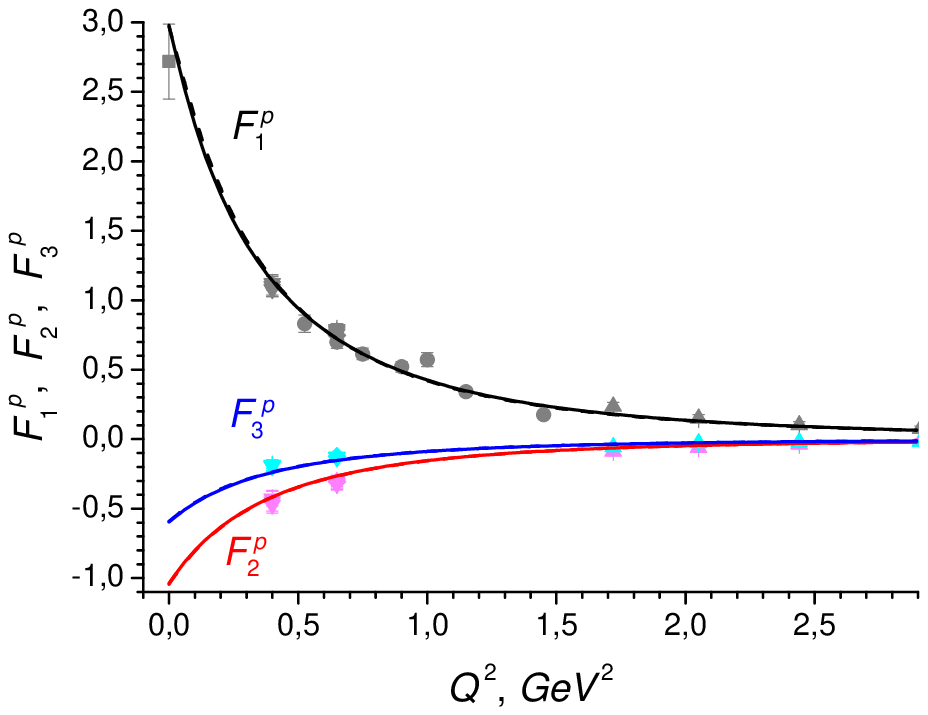}
	\includegraphics[width=0.49\linewidth]{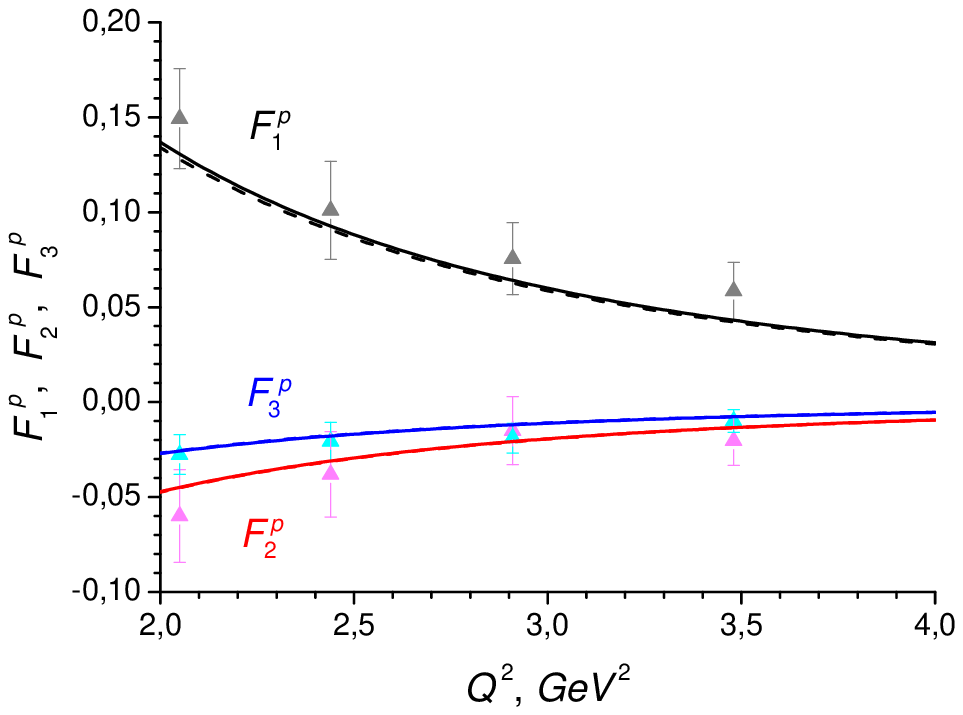}
	\caption{\label{fig:N(1520)FFF}Form factors of the model (\ref{G32}) extracted from the data on the transition $\gamma^*p\to N(1520)$. Curves and data are the same as in Fig.~\ref{fig:N(1520)AAS}.}
\end{figure*}

Available experimental data \cite{un-06,fr-99,ka-01,st-06,st-98,st-93,PDG,ti-04,az-05a,sp-05,el-06,ke-05} on the $\gamma^*N \to \Delta(1232)$ transition are depicted in Figs.~\ref{fig:Delta(1232)AAS}, \ref{fig:Delta(1232)REMRSM}, \ref{fig:Delta(1232)GM}. The extracted Lagrangian form factors $F_\alpha (Q^2)$ are pictured in Fig.~\ref{fig:Delta(1232)FFF}. The Fig.~\ref{fig:N(1520)AAS} presents the set of experimental data on the $\gamma^*N\to N(1520)$ amplitudes \cite{ti-04,az-05a,az-05b,la-06,PDG}. The form factors extracted from this data are in Fig.~\ref{fig:N(1520)FFF}. The fit of the amplitudes (\ref{A32})--(\ref{S12}) to the experimental data (the solid and dashed lines in all figures) is carried out in the framework of the QCD-inspired VMD model (see Secs. \ref{MVDM section}, \ref{fitsection}).

%==============================================================================================
%==============================================================================================
%==============================================================================================

\subsubsection{Quark-hadron duality and high-$Q^2$ form-factor behavior} \label{QHD}

The phenomenological model to interpret experimental data should obey the general implications of the quark-hadron duality. It makes the amplitudes $A_{1/2}(Q^2)$, $A_{3/2}(Q^2)$, $\tilde S_{1/2}(Q^2)$ and the form factors $F_\alpha(Q^2)$ to take on some specific properties in accordance with the origin of the form factors on both quark and hadron levels.

At very high momentum transfer, pQCD predicts the scaling behavior of the photoabsorption amplitudes to be \cite{ca-86}
\begin{equation}\label{AAS}
\begin{gathered}
	A_{1/2}(Q^2) \simeq \frac{C_{A(1/2)}}{Q^3}, \qquad A_{3/2}(Q^2) \simeq \frac{C_{A(3/2)}}{Q^5}, \\
	\tilde S_{1/2}(Q^2) \simeq \frac{C_S}{Q^4},
\end{gathered}
\end{equation}
where $C_{A(1/2)}$, $C_{S}$, $C_{A(3/2)}$ are constants or slight logarithmic functions of $Q^2$. This brings up the question of whether it is possible to obtain the power asymptotes of the form factors from those of the amplitudes (\ref{AAS}). The analysis of Eqs. (\ref{A32})--(\ref{S12}) and (\ref{F1f})--(\ref{F3f}) shows that only two form factors of the vertex (\ref{G32}) have uniquely determined asymptotes but the exponent of the third form factor asymptote is bounded below:
\begin{equation}\label{FFF}
\begin{gathered}
	F_{1}(Q^2) \simeq \frac{C_1}{Q^6}, \qquad F_{2}(Q^2)\simeq \frac{C_2}{Q^8}, \\
	F_{3}(Q^2)\simeq \frac{C_3}{Q^{2p_3}},\quad p_3\geqslant 4.
\end{gathered}
\end{equation}
To fit the experimental data, we suppose that $F_3(Q^2)\sim Q^{-8}$ for $Q^2\to \infty$.

In what follows we will exploit the three aspects of quark-hadron duality: 

\begin{enumerate}
\item In the asymptotic region of $Q^2\gg M_R^2$ resonance electroproduction is described in the  framework of QCD by only two independent form factors~--- the non-spin-flip $F_1(Q^2)$ and the spin-flip $F_2(Q^2)$. \textit{For $Q^2 \to \infty$ the transverse helicity amplitudes are proportional to different form factors while the ratio of their asymptotes is}
\begin{equation}
\displaystyle \frac{A_{3/2}(Q^2)}{A_{1/2}(Q^2)}\sim\frac{F_2(Q^2)}{F_1(Q^2)}\sim \frac{1}{Q^2}.
\label{AAFF}
\end{equation}
\item \textit{The asymptotic constrains must be imposed on the form factors so that the asymptotic scaling relation $R_\text{SM} \to const$ were valid.}
\item \textit{The longitudinal $\tilde S_{1/2}(Q^2)$ and transverse $A_{1/2}(Q^2)$ amplitudes are proportional to the same form factor $F_1(Q^2)$ at high $Q^2$.}
\end{enumerate}

The first two statements are due to the baryon helicity conservation at high $Q^2$. The third one arises from the fact that the absorption of a longitudinally polarized photon is asymptoticly a non-spin-flip interaction.

Taking all above considerations into account, one can easily obtain the asymptotes of the photoabsorption amplitudes (\ref{A32})--(\ref{S12}):
\begin{eqnarray}
	A_{1/2}(Q^2) &=& \displaystyle -\frac{1}{\sqrt{3}} N \frac{M_N}{M_R}Q^3F_1(Q^2), \label{A12ass}\\
	A_{3/2}(Q^2) &=& \displaystyle \pm \frac12 N Q^3F_2(Q^2), \label{A32ass}\\
	\tilde S_{1/2}(Q^2) &=& \displaystyle \pm N M_N Q^2 F_1(Q^2), \label{S12ass}
\end{eqnarray}
where $N = \displaystyle\left[\frac{\pi\alpha}{M_N^5(M_R^2-M_N^2)}\right]^{1/2}$.

In Eqs. (\ref{A12ass})--(\ref{S12ass}) it is assumed that the following inequalities hold for $Q^2\to \infty$:
\begin{equation}
\begin{array}{c}
\displaystyle F_1(Q^2)\gg\frac{M_R}{2M_N}\left|F_2(Q^2)-F_3(Q^2)\right|,
\\[5mm]
\displaystyle \left|F_2(Q^2)-F_3(Q^2)\right|\gg \frac{2M_N(M_R\pm
M_N)}{Q^2}\left|F_1(Q^2)\right|,
\\[5mm]
\displaystyle F_1(Q^2)\gg\left|\frac{M_R}{M_N}F_2(Q^2)+\frac{Q^2}{M_R
M_N}F_3(Q^2)\right|.
\end{array}
\label{gg}
\end{equation}

The first of the inequalities (\ref{gg}) is true by virtue of the form-factors asymptotes (\ref{FFF}), derived from the asymptotic pQCD-predictions (\ref{AAS}). But the last two of the inequalities (\ref{gg}) are valid only with regard to logarithmic renormalization. Note that logarithmic renormalization is inescapable due to the following reasons. There are at least two chromodynamic quark subprocesses contributing to the resonance electroproduction in the asymptotic region: (1) a single-quark transition to an excited state; (2) 4-momentum exchange between valence quarks. If these processes are short-distance and non-spin-flip, the amplitude $A_{1/2}(Q^2)$ is proportional to the third power of the strong coupling constant. Granting this consideration and the asymptotic relation $A_{1/2}(Q^2)\sim Q^3F_1(Q^2)$, it is easily seen that the non-spin-flip transition form factors obey quark counting rule
\begin{gather}\label{Fln}
	F_1(Q^2) \sim
	\left[ \frac{\alpha_s(Q^2)}{Q^2}\right]^{n_1}\sim \frac{1}{Q^{2n_1}\ln^{n_1}Q^2/\Lambda^2}, \nonumber \\
	n_1=n_{val}-1+n_{ex}=3,
\end{gather}
where $n_{val}=3$ is the number of the valence quarks; $n_{ex}=1$ is the number of the excited quarks; $\Lambda \approx \Lambda_\text{QCD}=0.215\pm 0.025 \text{ GeV}$ is the QCD scale parameter \footnote{If $n_{ex}=0$, Eq. (\ref{Fln}) turns into quark counting rule for the elastic Dirac form factor of the proton.}.

Modifying the asymptotes (\ref{FFF}) with the small parameter $\ln^{-1}Q^2/\Lambda^2\ll 1$, one readily imposes them to satisfy all the inequalities (\ref{gg}):
\begin{equation}
\displaystyle F_\alpha(Q^2) \simeq \left(\frac{4M_N^2}{Q^2}\right)^{p_\alpha}\cdot\frac{f_{\alpha}}{\ln^{n_\alpha}Q^2/\Lambda^2},
\label{FQCD}
\end{equation}
where $p_1=3$, $p_2= p_3= 4$, $n_3 > n_1 > n_2$, $n_1\simeq 3$, $f_\alpha$ are dimensionless parameters.

The asymptotic relation among the spin-flip and non-spin-flip form factors
\begin{equation}
\displaystyle \frac{F_2(Q^2)}{F_1(Q^2)} \sim \frac{\ln^nQ^2/\Lambda^2}{Q^2}, \qquad n=n_1-n_2,
\label{F2F1}
\end{equation}
resembles that among the elastic Pauli and Dirac form factor \footnote{In the case of the elastic $eN$-scattering the parameter $n \approx 2$ is exactly calculated in pQCD \cite{br-04}. However, the similar calculations of the asymptotic transition form factors have not been carried out yet. But it would appear reasonable to suppose by analogy to the elastic scattering that the logarithmically small parameter is $\ln^{-2} Q^2/\Lambda^2\ll 1$ for $n_1=3$ (this value is due to quark counting (\ref{Fln})) and $n_2=1$, $n_3=4$.} \cite{br-04}.

%==============================================================================================
%==============================================================================================
%==============================================================================================

\begin{widetext}
\subsubsection{The $P_{11}(1440)$, $S_{11}(1535)$}

The amplitudes for the electroproduction of spin-$1/2$ resonances calculated within the model (\ref{G12}) are
\begin{equation}
 \begin{array}{c}
 \displaystyle A_{1/2}(Q^2)=\sqrt{2}
  \left[\frac{\pi\alpha(Q^2+(M_R \mp M_N)^2)}{M_N^5(M_R^2-M_N^2)}\right]^{1/2}
 \left[Q^2G_1(Q^2)+M_N\left(M_R\pm M_N\right)G_2(Q^2)
  \right],
\end{array}
 \label{A1212}
 \end{equation}
 \begin{equation}
 \begin{array}{c}
 \displaystyle \tilde S_{1/2}(Q^2)\equiv
 S_{1/2}(Q^2)\left[1+\frac{(Q^2+M_R^2-M_N^2)^2}{4M_N^2Q^2}\right]^{-1/2}=
 \\[5mm]
 \displaystyle
 =\mp\left[\frac{\pi\alpha(Q^2+(M_R\mp M_N)^2)}
 {M_N^5(M_R^2-M_N^2)}\right]^{1/2}
 \left[(M_R\pm M_N) G_1(Q^2)-M_N G_2(Q^2)\right]Q.
 \end{array}
 \label{S1212}
 \end{equation}
In Eqs. (\ref{A1212}) and (\ref{S1212}) the top signs correspond to the $N(1440)$, while the bottom ones are for the $N(1535)$;  $G_1(Q^2)$ is the non-spin-flip form factor; $G_2(Q^2)$ is the spin-flip form factor.

The form factors extracted from the experimental data on the helicity amplitudes are
\begin{align} \label{G1}
	G_1(Q^2)= {}&
	\left[\frac{M_N^5(M_R^2-M_N^2)}{\pi\alpha[Q^2+(M_R\mp M_N)^2]}\right]^{1/2} \times {}
	\nonumber \\ &{}\times
	\frac{1}{[Q^2+(M_R\pm M_N)^2]}
	\left[\frac{ A_{1/2}(Q^2)}{\sqrt{2}}\mp\frac{M_R\pm M_N}{Q}\tilde S_{1/2}(Q^2)\right],
\end{align}
\begin{align} \label{G2}
	G_2(Q^2)= {}&
	\left[\frac{M_N^5(M_R^2-M_N^2)}{\pi\alpha[Q^2+(M_R\mp M_N)^2]}\right]^{1/2} \times {}
	\nonumber \\ &{}\times
	\frac{1}{[Q^2+(M_R\pm M_N)^2]}
	\left[\frac{M_R\pm M_N}{M_N}\frac{ A_{1/2}(Q^2)}{\sqrt{2}}\pm\frac{Q}{M_N}\tilde S_{1/2}(Q^2) \right],
\end{align}
\end{widetext}

\begin{figure*}
	\includegraphics[width=0.49\linewidth]{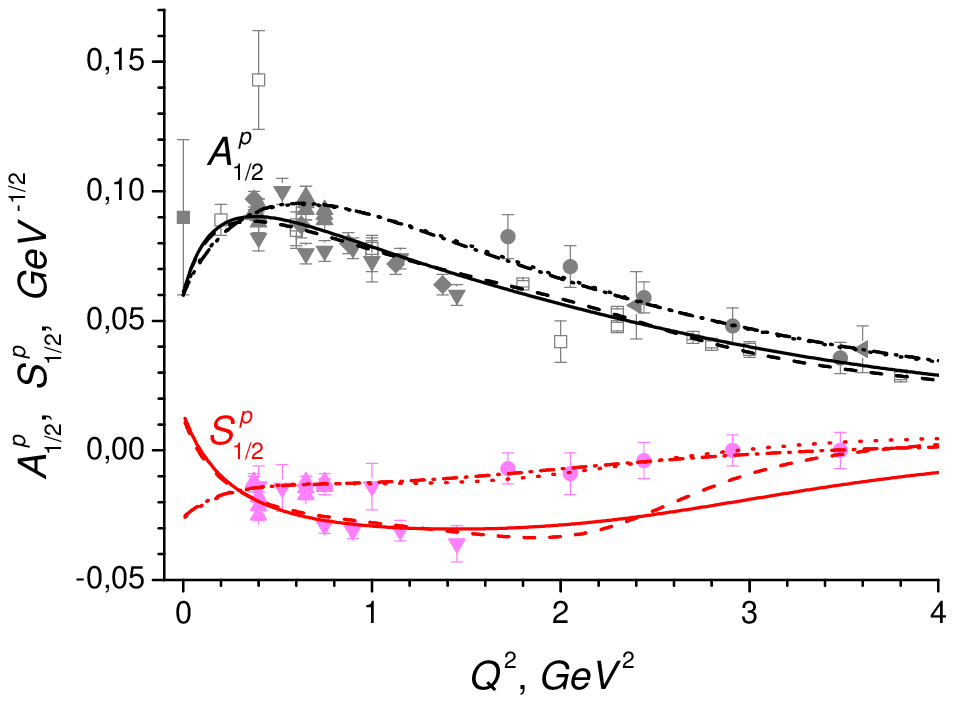}
	\includegraphics[width=0.49\linewidth]{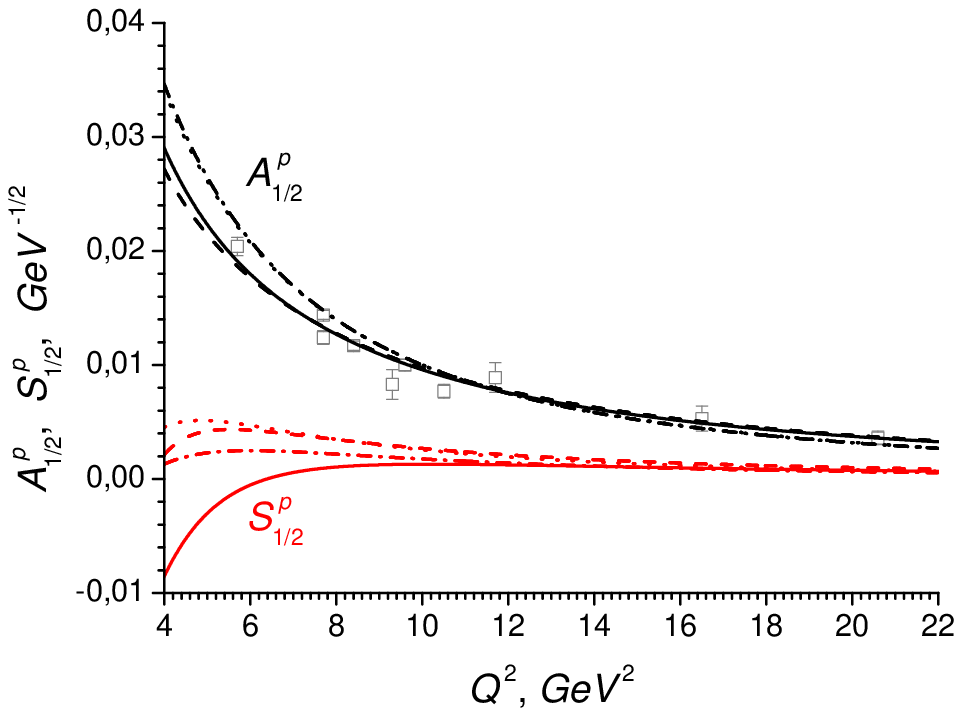}
	\caption{\label{fig:N(1535)AS}Helicity amplitudes of the transition $\gamma^*N \to N(1535)$. The solid curves correspond to fit S1-F1, dashed curves to S1-F2, dot-dashed curves to S2-F1, dotted curves to S2-F2 (see Sec.~\ref{The N(1535)}). The data points are denoted as follows: $\blacklozenge$ \cite{th-01}, {\large \textbullet} \cite{la-06}, {\footnotesize $\blacksquare$} \cite{PDG}, {\footnotesize $\square$} \cite{st-93}, $\blacktriangleleft$ \cite{ar-99}, $\blacktriangle$ \cite{az-05a}, $\blacktriangledown$ \cite{ti-04}.}
\end{figure*}

\begin{figure*}
	\includegraphics[width=0.49\linewidth]{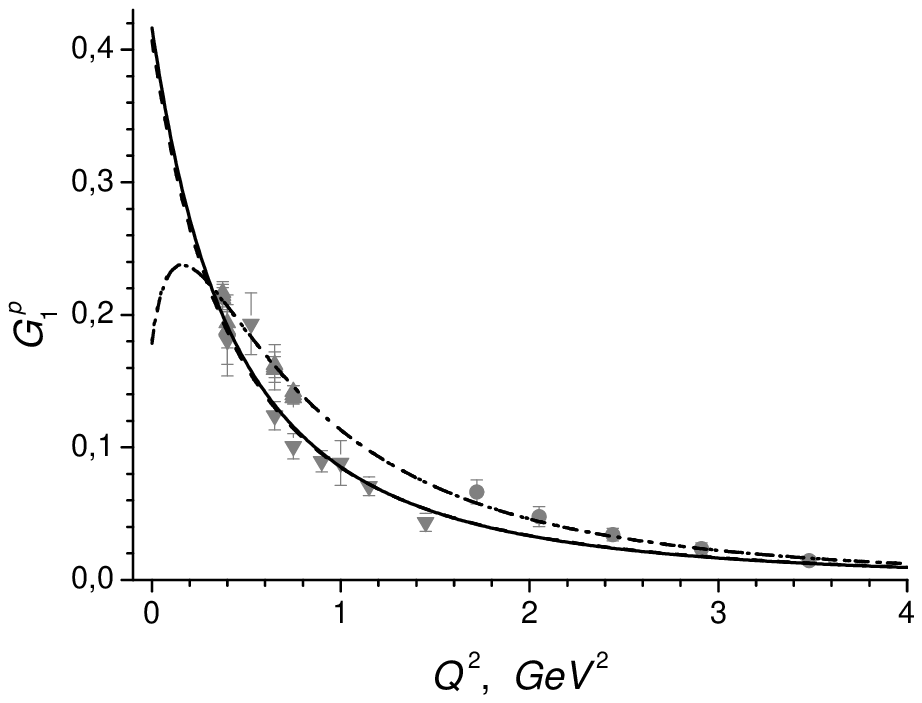}
	\includegraphics[width=0.49\linewidth]{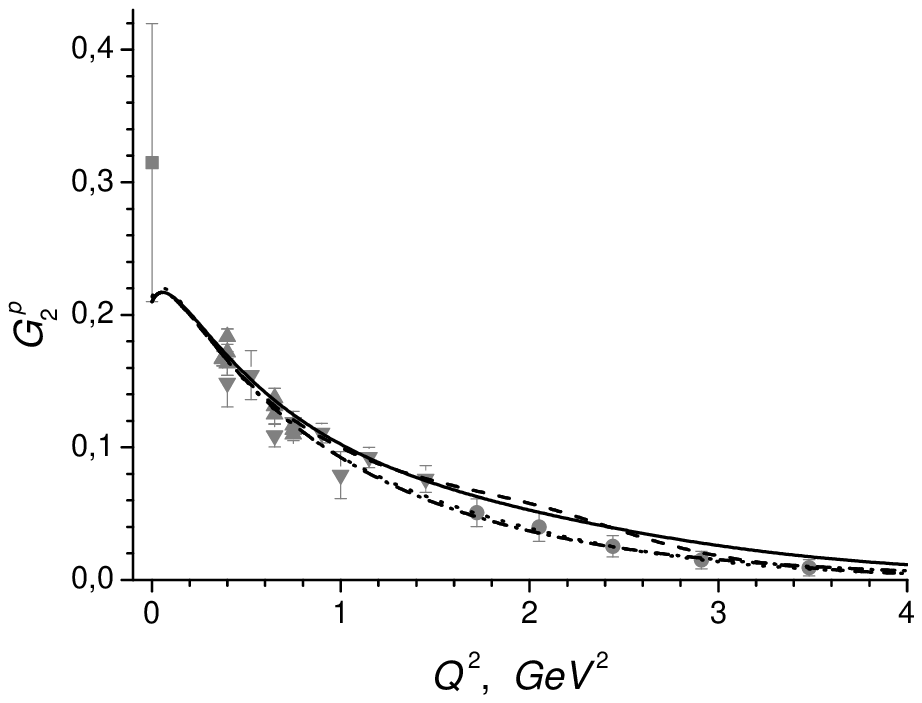}
	\caption{\label{fig:N(1535)GG}Form factors of the model (\ref{G12}) extracted from the data on the transition $\gamma^*N \to N(1535)$.  Curves and data are the same as in Fig.~\ref{fig:N(1535)AS}.}
\end{figure*}

\begin{figure*}
	\includegraphics[width=0.49\linewidth]{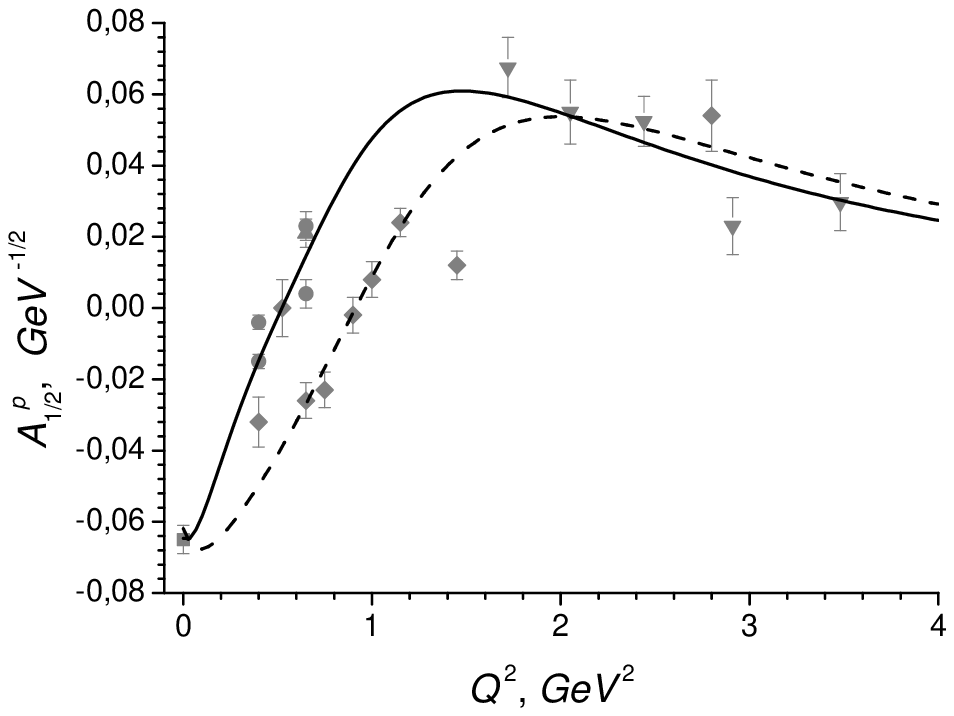}
	\includegraphics[width=0.49\linewidth]{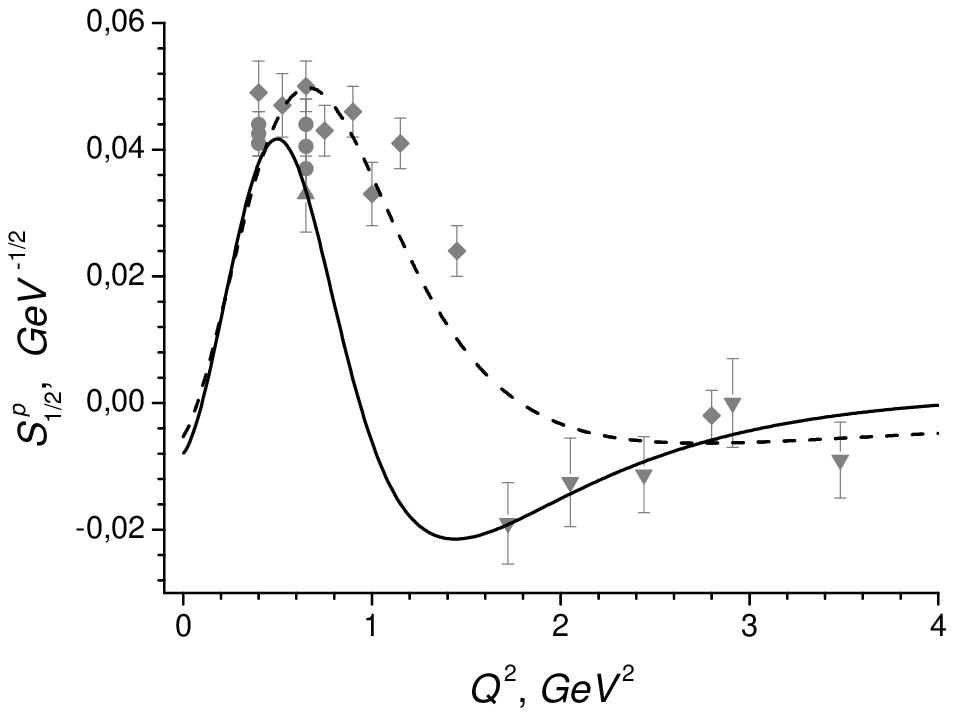}
	\caption{\label{fig:N(1440)AS}Helicity amplitudes of the transition $\gamma^*N \to N(1440)$. The solid curves correspond to the fit to the data sample S1, dashed curves to S2 (see Sec.~\ref{The N(1440)}). The data points are denoted as follows: $\blacklozenge$ \cite{ti-04}, {\large \textbullet} \cite{az-05a}, {\footnotesize $\blacksquare$} \cite{PDG}, $\blacktriangle$ \cite{az-05b}, $\blacktriangledown$ \cite{la-06}.}
\end{figure*}

\begin{figure*}
	\includegraphics[width=0.49\linewidth]{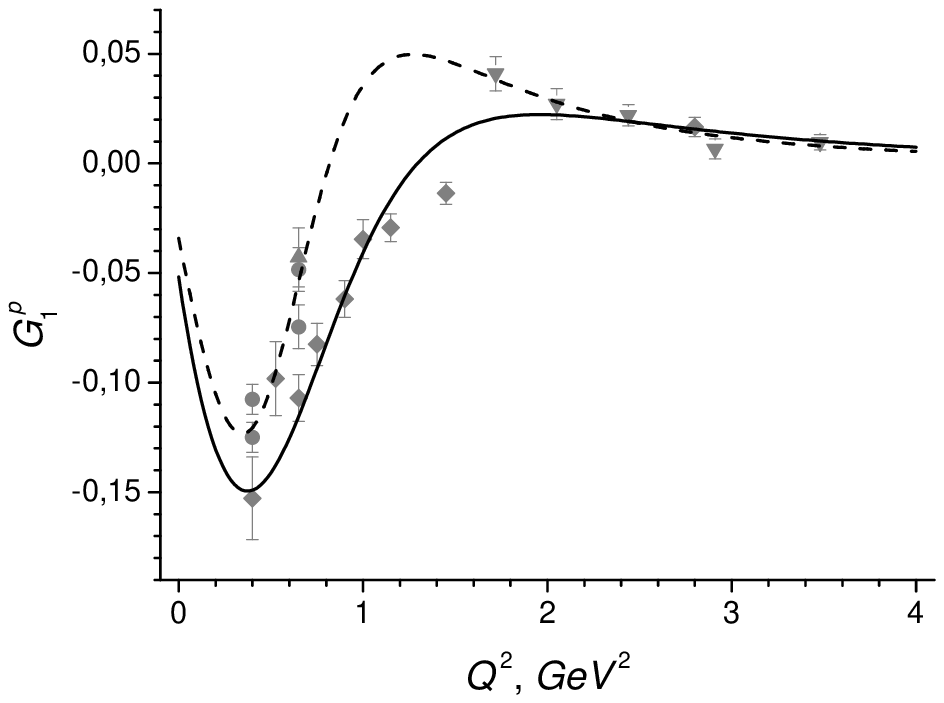}
	\includegraphics[width=0.49\linewidth]{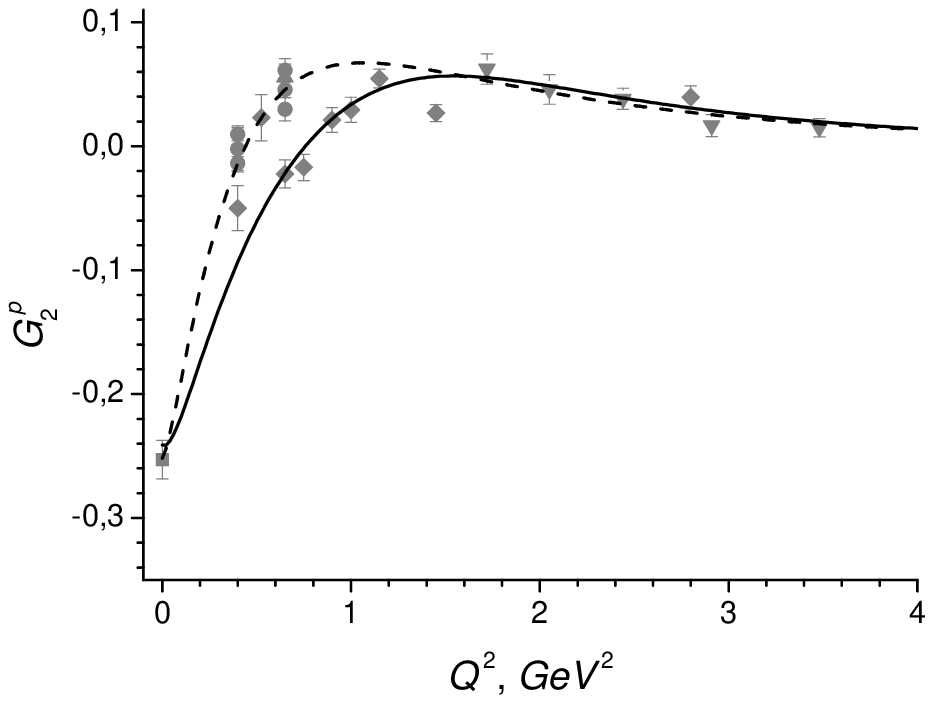}
	\caption{\label{fig:N(1440)GG}Form factors of the model (\ref{G12}) extracted from the data on the transition $\gamma^*N \to N(1440)$.  Curves and data are the same as in Fig.~\ref{fig:N(1440)AS}.}
\end{figure*}

The experimental data on the photoabsorption amplitudes \cite{st-93,az-05a,az-05b,ti-04,ar-99,th-01,PDG} and the extracted Lagrangian form factors are pictured in Figs.~\ref{fig:N(1440)AS} and \ref{fig:N(1440)GG} for the transfer $\gamma^*p \to N(1440)$ and in Figs.~\ref{fig:N(1535)AS} and \ref{fig:N(1535)GG} for the $\gamma^*p \to N(1535)$.

The asymptotic behavior of the helicity amplitudes predicted by pQCD are
\begin{equation}\label{AS12ass}
	A_{1/2}(Q^2) \simeq \frac{C_A}{Q^3}, \quad
	\tilde S_{1/2}(Q^2) \simeq \frac{C_S}{Q^4}.
\end{equation}
At high $Q^2$ the dominant contribution to both the electromagnetic amplitudes is to be come from only the non-spin-flip interactions. Granting this and substituting (\ref{AS12ass}) into (\ref{A1212}), one can easily obtain the asymptotic behavior of the first form factor and the limitations on the asymptote of the second form factor:
\begin{equation}\label{G1ass}
	G_1(Q^2) \simeq  \frac{C_1}{Q^6}, \quad
	G_2(Q^2) \simeq  \frac{C_2}{Q^{2p_2}}, \qquad
	p_2>3.
\end{equation}

Logarithmic renormalization of the form factor $G_1(Q^2)$ can be carried out with regard to the same quark counting rule as \eqref{Fln}. Besides, it is reasonable to suggest the ratio $G_2(Q^2)/G_1(Q^2)$ at high $Q^2$ to be the same as in the case of spin-vector resonance electroproduction (\ref{F2F1}). In the framework of the two latter considerations, one can easily obtain the following form-factor asymptotes for spin-$1/2$ resonances:
\begin{equation}
\begin{array}{c}
\displaystyle
 G_\alpha(Q^2)\simeq
\left(\frac{4M_N^2}{Q^2}\right)^{p_\alpha}\cdot\frac{g_{\alpha}}{\ln^{n_\alpha}Q^2/\Lambda^2},
\\[5mm]
p_1=3,\quad p_2= 4;\qquad\qquad n_1 > n_2,\quad n_1\simeq 3.
\end{array}
\label{GQCD}
\end{equation}

%==============================================================================================
%==============================================================================================
%==============================================================================================

\subsection{Electromagnetic coupling constants}

By definition, the electromagnetic constants of the resonances are the amplitudes for the absorption of a real transverse photon $A_{1/2}(0)$, $A_{3/2}(0)$. It is also possible to define model  electromagnetic constants, i.e., the Lagrangian form factors at $Q^2=0$. Some of such parameters within the models (\ref{G32}) and (\ref{G12}) are
\begin{widetext}
\begin{equation}
\begin{array}{c}
\displaystyle F_1(0)=\left[\frac{M_N^5}{\pi\alpha(M_R^2-M_N^2)}\right]^{1/2}
\frac{M_R}{M_N(M_R\pm M_N)}(\mp A_{3/2}-\sqrt{3}A_{1/2})\, ,
\\[5mm]
\displaystyle
F_2(0)+F_3(0)=\left[\frac{M_N^5}{\pi\alpha(M_R^2-M_N^2)}\right]^{1/2}
\frac{2}{M_R^2- M_N^2}(-M_N A_{3/2}+M_R\sqrt{3}A_{1/2})\, ,
\\[5mm]
\displaystyle
G_2(0)=\left[\frac{M_N^5}{\pi\alpha(M_R^2-M_N^2)}\right]^{1/2}
\frac{1}{\sqrt{2}M_N}A_{1/2}\, .
\end{array}
\label{FG0}
\end{equation}
The form factors and observed values of the amplitudes at $Q^2=0$~\cite{PDG} are set out in Table~\ref{tab:LEC}.
\newpage
\end{widetext}

\begin{table*}
\caption{\label{tab:LEC}Electromagnetic coupling constants~\cite{PDG}.}
\begin{ruledtabular}
\tabcolsep6.8pt
\renewcommand{\arraystretch}{1.3}
\begin{tabular}{lcccc}
$A_h(0),\,F_\alpha(0),\,G_\alpha(0)$ & $P_{33}\,\Delta(1232)$ & $D_{13}\, N(1520)$& $P_{11}\, N(1440)$& $S_{11}\, N(1535)$ \\
\hline
$A_{1/2}^{(p)}(0), \text{ GeV}^{-1/2}$ & $-0.135 \pm 0.006$ & $-0.024 \pm 0.009$ & $-0.065 \pm 0.004$ & $\phm 0.090 \pm 0.030$ \\
$A_{3/2}^{(p)}(0), \text{ GeV}^{-1/2}$ & $-0.25 \phn \pm 0.008 $ & $\phm 0.166 \pm 0.005$ & --- & --- \\
\hline
$F_{1}^{(p)}(0)$ & $\phm 2.068\pm 0.079$ & $\phm 2.729 \pm 0.271$ & --- & --- \\
$F_2^{(p)}(0)+F_3^{(p)}(0)$ & $-1.186 \pm 0.452$ & $-1.447 \pm 0.188$ & --- & --- \\
$G_{2}^{(p)}(0)$ & --- & --- & $-0.253 \pm 0.016$ & $\phm 0.315 \pm 0.105$ \\
\hline\hline
$A_{1/2}^{(n)}(0), \text{ GeV}^{-1/2}$ & $\phm 0.135 \pm 0.006$ & $-0.059 \pm 0.009$ & $\phm 0.040 \pm 0.010$ & $-0.046 \pm 0.027$ \\
$A_{3/2}^{(n)}(0), \text{ GeV}^{-1/2}$ & $\phm 0.25 \phn \pm 0.008$ & $-0.139 \pm 0.011$ & --- & --- \\
\hline
$F_{1}^{(n)}(0)$ & $-2.068 \mp 0.079$ & $-0.484 \pm 0.350$ & --- & --- \\
$F_2^{(n)}(0)+F_3^{(n)}(0)$ & $\phm 1.186 \pm 0.452$ & $-0.164 \pm 0.225$ & --- & --- \\
$G_{2}^{(n)}(0)$ & --- & --- & $\phm 0.156 \pm 0.039$ & $-0.161 \pm 0.094$ \\
\end{tabular}
\end{ruledtabular}
\end{table*}

%==============================================================================================
%==============================================================================================
%==============================================================================================

\section{Form factors within VMD model} \label{MVDM section}

%==============================================================================================
%==============================================================================================
%==============================================================================================

\subsection{Origin of the model}

Vector-meson-dominance models being consistent with pQCD-predictions are well known to give a satisfactory description of existing experimental data on elastic $eN$-scattering \cite{gk-85,gk-92,ia-04a,ia-04b,ia-04c,lo-01,lo-02,du-02}. The universal physical ground of VMD allows to apply its principles to physics of the transition form factors. But the VMD models currently in use are suffering the drawback of taking into account solely the ground-state vector mesons $\rho(770)$,  $\omega(782)$, $\phi(1020)$ \cite{gk-85,gk-92,ia-04a,ia-04b,ia-04c} and seldom $\rho(1450)$ \cite{lo-01,lo-02} and $\omega(1420)$ \cite{lo-02}. This cut-off of the meson spectrum is usually motivated by the data on decay widths $\Gamma(V \to e^-e^+)$ testifying a photon to hadronize dominantly into the above mesons \cite{gk-85}. To join predictions of such VMD models with pQCD expectations, the hadronization amplitudes should be suppressed by power and logarithmic functions.

However, the truncation of the intermediary vector mesons spectrum and suppression of the amplitudes by artificial means are in conflict with physics of the process and beyond the framework of quantum field theory. Actually, in the nonperturbative hadronic vacuum a photon excites all  modes of hadronic string, carrying the quantum numbers $J^{PC}=1^{--}$. Thus, all the vector mesons (at least the observed ones) should be incorporated in the VMD model. Furthermore, the low values of the hadronization amplitudes is not an adequate cause to disregard of heavy mesons. This is due to the structure of the amplitudes for the transition $eN \to eR$:
\begin{align}\label{AV}
\mathcal{A}_{V(k)}(\gamma^*N&\to R)=
\sum_\alpha\mathcal{A}_{\alpha k}^{(V)}(\gamma^*N\to R) = 
\nonumber \\
&\mathcal{A}(\gamma^* \to V^*_k) \times \sum_\alpha\mathcal{A}_\alpha(V^*_kN \to R),
\end{align}
where $ \mathcal{A}(\gamma^* \to V^*_k) $ is the amplitude for the transition of a virtual photon to virtual vector meson; $\mathcal{A}_\alpha(V^*_kN \to R) $ is the amplitude for the absorption of a virtual meson by nucleon; the ``$\alpha$'' indexes the vertexes of meson-nucleon coupling corresponding to independent form factors. In the case of high excited resonances, the photoabsorption amplitudes $\mathcal{A}_{\alpha k}^{(V)}(\gamma^*N\to R)$ are not necessarily negligible since small hadronization amplitudes are multiplied by arbitrary large meson-absorption amplitudes. Note that it is the set of the amplitudes $\mathcal{A}_{\alpha k}^{(V)}(\gamma^*N\to R)$ [not $\mathcal{A}_\alpha(V^*_kN \to R)$] that is obtained by the fit to experimental data [the meson-absorption amplitudes can be then easily calculated if the $\mathcal{A}_{\alpha k}^{(V)}(\gamma^*N\to R)$ are known].

Light unflavored mesons listed in Particle Data Group tables \cite{PDG} are grouped by near mass degeneracy into five singlet-triplet families (see Table~\ref{tab:mesons}).

\begin{table}
\caption{\label{tab:mesons}PDG vector-meson masses.}
\begin{ruledtabular}
\begin{tabular}{cccccc}
$k$&$$&$m_{(\rho)k},\text{ GeV}$&&$m_{(\omega)k},\text{ GeV}$&$m_k,\text{ GeV}$\footnotemark[1]\\
\hline
$1$ & $\rho(770)\hphantom{0}$ & 0.7755 & $\omega(782)\hphantom{0\footnotemark[2]}$ & 0.78265 & $0.7791$ \\
$2$ & $\rho(1450)$ & 1.459\hphantom{0}& $\omega(1420)\hphantom{\footnotemark[2]}$ & 1.425\hphantom{00}& $1.442\hphantom{0}$ \\
$3$ & $\rho(1700)$ & 1.720\hphantom{0}& $\omega(1650)\hphantom{\footnotemark[2]}$ & 1.670\hphantom{00}& $1.695\hphantom{0}$ \\
$4$ & $\rho(1900)$ & 1.885\hphantom{0}& $\omega(1960)$\footnotemark[2] & 1.960\hphantom{00}& $1.923\hphantom{0}$ \\
$5$ & $\rho(2150)$ & 2.149\hphantom{0}& $\omega(2145)$\footnotemark[2] & 2.148\hphantom{00}& $2.149\hphantom{0}$ \\
\end{tabular}
\end{ruledtabular}
\footnotetext[1]{$m_k = \left[(m_{(\rho)k}^2+m_{(\omega)k}^2)/2\right]^{1/2}$ is an averaged mass used in the fits for the second-region resonances (see Sec.~\ref{The second resonance region}).}
\footnotetext[2]{These isosinglet mesons are from ``Further states'' section.}
\end{table}

In the general case, $\phi${\nobreakdash-\hspace{0pt}}mesons are other intermediaries in $eN${\nobreakdash-\hspace{0pt}}interactions. However, to simplify the VMD model, we neglect their contribution to transition form factors due to the following reasons. In the case of ideal singlet-octet mixing corresponding to the quark content $\phi = \bar ss$, these mesons interact only with the strange component of the nucleon which is suppressed with respect to nonstrange quark content. The difference between actual and ideal mixing is also suppressed by small parameters, and to the first approximation in these parameters it is possible not to take into account coupling between $\phi${\nobreakdash-\hspace{0pt}}mesons and $ud${\nobreakdash-\hspace{0pt}}component of the nucleon.

So, to the extent that $\phi${\nobreakdash-\hspace{0pt}}mesons contributions can be neglected, the transition form factors are specified by dispersionlike expansions with poles at meson masses. The expansion coefficients are the amplitudes $\mathcal{A}_{\alpha k}^{(V)}(\gamma^*N\to R)$ [in the following we use this designation only for $R = \Delta(1232),\; N(1520)$ and $\mathcal{B}_{\alpha k}^{(V)}(\gamma^*N \to R)$ for $R = N(1440),\; N(1535)$]. Having regard for the isotopic symmetry of strong interactions, all the transition form factors are given by the sum over isosinglet and isovector contributions:
\begin{widetext}
\begin{align}
	&\Delta(1232),\; N(1520):&
	&F^{(p,n)}_\alpha(Q^2)=
	\frac12\sum_{k=1}^K\left[\mathcal{A}^{(\omega)}_{\alpha k}(Q^2)\frac{m_{(\omega)k}^2}{Q^2+m_{(\omega)k}^2}
	\pm
	\mathcal{A}^{(\rho)}_{\alpha k}(Q^2)\frac{m_{(\rho)k}^2}{Q^2+m_{(\rho)k}^2}\right],
	\label{Fvdm}
	\\
	&N(1440),\; N(1535):&
	&G^{(p,n)}_\alpha(Q^2)=
	\frac12\sum_{k=1}^K\left[\mathcal{B}^{(\omega)}_{\alpha k}(Q^2)\frac{m_{(\omega)k}^2}{Q^2+m_{(\omega)k}^2}
	\pm
	\mathcal{B}^{(\rho)}_{\alpha k}(Q^2)\frac{m_{(\rho)k}^2}{Q^2+m_{(\rho)k}^2}\right].
	\label{Gvdm}
\end{align}
\end{widetext}
Because of the value of the $\Delta(1232)$ isospin, $\rho$-mesons are only intermediaries in the $N\Delta$-coupling, i.e., $\mathcal{A}_{\alpha k}^{(\omega)}(Q^2)=0$.

Dispersionlike expansions of the form factors are predicted by the foundations of quantum field theory, that are taken into account by the dispersion relation approach. In the one-meson exchange approximation and in the limit of narrow-width mesons, the expansion coefficients $\mathcal{A}_{\alpha k}^{(V)}(Q^2)$, $\mathcal{B}_{\alpha k}^{(V)}(Q^2)$ are constants. But in Eqs. (\ref{Fvdm}) and (\ref{Gvdm}) they are supposed to be logarithmic functions of $Q^2$. It has been pointed out above in the Sec. \ref{QHD} that logarithmic renormalization of the form factors is demanded by quark-hadron duality. To this must be added that the logarithmic renormalization is also imposed by short-distance quark-gluon processes influencing the photon transition to mesons inside nucleon, i.e., at $Q^2>R_N^{-2}=(0.2\,\text{ GeV})^2$. Logarithmic factors at expansion coefficients absorb in phenomenological fashion the effects of the renormalization of the strong coupling constant and $Q^2${\nobreakdash-\hspace{0pt}}evolution of the parton distribution functions.

%==============================================================================================
%==============================================================================================
%==============================================================================================

\subsection{Asymptotic behavior of the dispersionlike expansions}

At high $Q^2$ the form factors \eqref{Fvdm} and \eqref{Gvdm} should join pQCD-predictions \eqref{FQCD} and \eqref{GQCD}. This property requires the expansion coefficients $\mathcal{A}_{\alpha k}^{(\omega,\,\rho)}(Q^2)$, $\mathcal{B}_{\alpha k}^{(\omega,\,\rho)}(Q^2)$ to obey a number of relations that we refer to from now on as the superconvergence relations (SRs).

Since logarithmic renormalization has a bearing to only the QCD effects taking place inside nucleon, it seems justified to suppose that logarithmic $Q^2${\nobreakdash-\hspace{0pt}}dependence of the form-factor expansion coefficients $\mathcal{A}_{\alpha k}^{(V)}$ is universal function $L_{\mathcal{A}\alpha}^{(V)}(Q^2)$ independent of meson family index $k=1,2,...,K$ (similarly $L_{\mathcal{B}\alpha}^{(V)}$ is $Q^2$-dependence of the amplitudes $\mathcal{B}_{\alpha k}^{(V)}$ up to numeric factors). Then, the isosiglet and isotriplet running electromagnetic coupling parameters can be represented in the following form:
\begin{equation}
\begin{array}{c}
\displaystyle
\sum_{k=1}^K\mathcal{A}_{\alpha k}^{(\omega,\,\rho)}(Q^2)\equiv\varkappa_\alpha^{(\omega,\,\rho)}(Q^2)=
\frac{\varkappa_\alpha^{(\omega,\,\rho)}(0)}{L_{\mathcal{A}\alpha}^{(\omega,\,\rho)}(Q^2)},
\\[5mm]\displaystyle
\sum_{k=1}^K\mathcal{B}_{\alpha k}^{(\omega,\,\rho)}(Q^2)\equiv\kappa_\alpha^{(\omega,\,\rho)}(Q^2)=
\frac{\kappa_\alpha^{(\omega,\,\rho)}(0)}{L_{\mathcal{B}\alpha}^{(\omega,\,\rho)}(Q^2)},
\end{array}
\label{kap}
\end{equation}
where
\begin{equation}\label{kaF}
	\begin{split}
	\varkappa_\alpha^{(\omega,\,\rho)}(0) &= F^{(p)}_\alpha(0)\pm F^{(n)}_\alpha(0),
	\\
	\kappa_\alpha^{(\omega,\,\rho)}(0) &= G^{(p)}_\alpha(0)\pm G^{(n)}_\alpha(0)
	\end{split}
\end{equation}
are the values of the parameters at $Q^2=0$. The logarithmic functions $L_{\mathcal{A}\alpha}^{(V)}(Q^2)$ and $L_{\mathcal{B}\alpha}^{(V)}(Q^2)$ are known in the static and asymptotic limit:
\begin{equation}\label{L}
L_{\alpha}^{(V)}(Q^2) \to \Biggl\{
\begin{aligned}
&1, &Q^2 &\to 0, \\
&C_\alpha^{(V)} \ln^{n_\alpha}{Q^2/\Lambda^2},&Q^2 &\to \infty.
\end{aligned}
\end{equation}
The most simple interpolation function retaining the asymptotic behavior (\ref{L}) is
\begin{equation}\label{Li1}
	\displaystyle L_{\alpha}^{(V)}(Q^2)=1+ C_{\alpha}^{(V)}\ln^{n_\alpha}\left(1+\frac{Q^2}{\Lambda^2}\right).
\end{equation}
Another possibility is
\begin{align}\label{Li2}
	L_{\alpha}^{(V)} (Q^2) = \biggl[&1+h_{\alpha}^{(V)}\ln\left(1+\frac{Q^2}{\Lambda^2}\right)+\nonumber\\
	&\displaystyle k_{\alpha}^{(V)}\ln^2\left(1+\frac{Q^2}{\Lambda^2}\right)\biggr]^{n_\alpha/2}.
\end{align}
The interpolation function \eqref{Li2} effectively takes into account effects of nonleading pQCD-logarithms.

The expansion coefficients are proportional to the running coupling parameters. The numeric dimensionless factors of proportionality are denoted as follows:
\begin{equation}\label{ab}
	\begin{array}{c}
	\displaystyle
	\frac{\mathcal{A}_{\alpha k}^{(V)}(Q^2)}{\varkappa_\alpha^{(V)}(Q^2)}
	=a_{\alpha k}^{(V)}=const,
	\quad \sum_{k=1}^Ka_{\alpha k}^{(V)}=1,
	\\[5mm]
	\displaystyle
	\frac{\mathcal{B}_{\alpha k}^{(V)}(Q^2)}{\kappa_\alpha^{(V)}(Q^2)}
	=b_{\alpha k}^{(V)}=const,
	\quad \sum_{k=1}^Kb_{\alpha k}^{(V)}=1.
	\end{array}
\end{equation}
Now the form factors (\ref{Fvdm}) and (\ref{Gvdm}) can be expressed in terms of the parameters introduced above:
\begin{widetext}
\begin{align}
	&\Delta(1232),\, N(1520): &
	&F^{(p,n)}_\alpha (Q^2) =
	\displaystyle \frac12 \Biggl[ \varkappa_\alpha^{(\omega)} (Q^2) \sum_{k=1}^K \frac{a_{\alpha k}^{(\omega)} m_{(\omega)k}^2}{Q^2+m_{(\omega)k}^2} \pm
	\varkappa_\alpha^{(\rho)} (Q^2) \sum_{k=1}^K \frac{a_{\alpha k}^{(\rho)} m_{(\rho)k}^2}{Q^2+m_{(\rho)k}^2} \Biggr],
	\label{Fvdm1}
	\\
	&N(1440),\, N(1535): &
	&G^{(p,n)}_\alpha (Q^2) =
	\displaystyle \frac12 \Biggl[ \kappa_\alpha^{(\omega)} (Q^2) \sum_{k=1}^K \frac{b_{\alpha k}^{(\omega)} m_{(\omega)k}^2}{Q^2+m_{(\omega)k}^2} \pm
	\kappa_\alpha^{(\rho)} (Q^2) \sum_{k=1}^K \frac{b_{\alpha k}^{(\rho)} m_{(\rho)k}^2}{Q^2+m_{(\rho)k}^2} \Biggr],
	\label{Gvdm1}
\end{align}
\end{widetext}
where $\varkappa_\alpha^{(\omega)} (Q^2) \equiv 0$ for the $\Delta(1232)$.
To assure the correct asymptotic behavior (\ref{FQCD}) and (\ref{GQCD}) of the form factors (\ref{Fvdm1}) and (\ref{Gvdm1}), one should expand them in powers of $1/Q^2$ and set the coefficients preceding $Q^{-2}$, $Q^{-4}$ (and $Q^{-6}$ in the case of $F_{2,3}$, $G_2$) equal to zero. The constraints obtained in the fashion described are the SRs between the meson parameters $a_{\alpha k}^{(V)}$, $b_{\alpha k}^{(V)}$.

The set of the SRs between the parameters of the transition $N\to\Delta(1232)$ are
\begin{align}\label{SR1}
	\alpha = {}&1: \phantom{,3}  \nonumber\\&
	\begin{aligned}
	& \sum_{k=1}^K a_{1k}^{(\rho)} = 1,
	& \sum_{k=1}^K a_{1k}^{(\rho)} m_{(\rho)k}^2 = 0, \\
	& \sum_{k=1}^K a_{1k}^{(\rho)} m_{(\rho)k}^4 = 0;
	\end{aligned}
\end{align}
\begin{align}\label{SR2}
	\alpha = {}&2,3: \phantom{,3}\nonumber\\&
	\begin{aligned}
	& \sum_{k=1}^K a_{\alpha k}^{(\rho)} = 1,
	& \sum_{k=1}^K a_{\alpha k}^{(\rho)} m_{(\rho)k}^2 = 0, \\
	& \sum_{k=1}^K a_{\alpha k}^{(\rho)} m_{(\rho)k}^4 = 0,
	& \sum_{k=1}^K a_{\alpha k}^{(\rho)} m_{(\rho)k}^6 = 0.
	\end{aligned}
\end{align}
In the case of the transition $N \to N(1520)$ the nonvanishing isosinglet contributions to the form factors lead to some more SRs between the parameters of $\omega${\nobreakdash-\hspace{0pt}}mesons:
\begin{align}\label{SR3}
	\alpha = {}&1:  \phantom{,3}\nonumber\\&
	\begin{aligned}
	& \sum_{k=1}^K a_{1k}^{(\omega)} = 1,
	& \sum_{k=1}^K a_{1k}^{(\omega)} m_{(\omega)k}^2 = 0, \\
	& \sum_{k=1}^K a_{1k}^{(\omega)} m_{(\omega)k}^4 = 0;
	\end{aligned}
\end{align}
\begin{align}\label{SR4}
	\alpha = {}&2,3: \phantom{,3}\nonumber\\&
	\begin{aligned}
	& \sum_{k=1}^K a_{\alpha k}^{(\omega)} = 1,
	& \sum_{k=1}^K a_{\alpha k}^{(\omega)} m_{(\omega)k}^2 = 0, \\
	& \sum_{k=1}^K a_{\alpha k}^{(\omega)} m_{(\omega)k}^4 = 0,
	& \sum_{k=1}^K a_{\alpha k}^{(\omega)} m_{(\omega)k}^6 = 0.
	\end{aligned}
\end{align}
The SRs similar to (\ref{SR1})--(\ref{SR4}) are valid for the parameters $b_{\alpha k}^{(\omega,\,\rho)}$ of the form factors for the $N \to N(1440)$ and $N \to N(1535)$ transitions.

%==============================================================================================
%==============================================================================================
%==============================================================================================

\section{Data analisis. Discussion and predictions}\label{fitsection}

%==============================================================================================
%==============================================================================================
%==============================================================================================

\subsection{The $\Delta(1232)$}

From the point of view of helicity-amplitude fitting, the $\Delta(1232)$ resonance offers an important simplification: its excitation via electroproduction off nucleon is only by photon and $\rho$-mesons, which halves the number of dispersionlike expansion coefficients. Besides, the data on the $\Delta(1232)$ helicity amplitudes is much more vast and precise compared to the data sets on other resonant amplitudes. All that allows the form factors $F_1(Q^2)$, $F_2(Q^2)$, $F_3(Q^2)$ to be extracted to a high accuracy.

In the case of the $\Delta(1232)$, the described VMD model gives the following expressions for the form factors 
\begin{eqnarray}\label{F1F2F3}
	\begin{aligned}
	\displaystyle F_1(Q^2) &= \frac{F^{(exp)}_1}{L_{1}^{(\rho)}(Q^2)}
	\sum_{k=1}^K \frac{a^{(\rho)}_{1k} m^2_{(\rho)k}}{Q^2+m^2_{(\rho)k}},
	\\
	\displaystyle F_2(Q^2) &= \frac{F_{2}(0)}{L_{2}^{(\rho)}(Q^2)}
	\sum_{k=1}^K \frac{a^{(\rho)}_{2k} m^2_{(\rho)k}}{Q^2+m^2_{(\rho)k}},
	\\
	\displaystyle F_3(Q^2) &= \frac{F^{(exp)}_{23}-F_{2}(0)}{L_{3}^{(\rho)}(Q^2)}
	\sum_{k=1}^K \frac{a^{(\rho)}_{3k} m^2_{(\rho)k}}{Q^2+m^2_{(\rho)k}},
	\end{aligned}
\end{eqnarray}
where $F^{(exp)}_{1}=2.068\pm 0.079$, $F^{(exp)}_{23}\equiv F_2(0)+F_3(0)=-1.186 \pm 0.452$ --- measured electrodynamic parameters; $L_{1,2,3}^{(\rho)}(Q^2)$ --- logarithmic functions satisfying the asymptotes (\ref{L}).

The number of the form-factor poles $K$ is bounded above by the cut-off of the $\rho$-meson spectrum and below by the number of the SRs \eqref{SR1} and \eqref{SR2}. In this simplest model $K=4$ dealing with only the first four $\rho$-mesons from Table~\ref{tab:mesons}, all the parameters $a^{(\rho)}_{\alpha k}$, $\alpha=2,3$ are fixed by the four SRs \eqref{SR2} and $a^{(\rho)}_{2k}=a^{(\rho)}_{3k}$. From this point on, we restrict the discussion to the specific case of the model with $K=4$, in order to reduce the number of free parameters. However, as it was pointed out in Sec. \ref{MVDM section}, there are no physical reasons to cut off the meson spectrum artificially. In fact, the spectrum should be truncated at highly excited vector states with widths exceeding inverse hadronization time $\Gamma_V > T_g^{-1}\simeq 1.2 \, \text{ GeV}$. Nevertheless, the inclusion of all the vector mesons is impossible, for it will overparametrize the fit.

%==============================================================================================
%==============================================================================================
%==============================================================================================

%==============================================================================================
%==============================================================================================
%==============================================================================================

\subsubsection{Fit results}

In the simplest model with $K=4$ incorporating only the first four $\rho$-mesons, all the parameters $a^{(\rho)}_{2k}$ are determined by the four SRs \eqref{SR2}. Also the parameters $a^{(\rho)}_{1k}$ satisfy three SRs \eqref{SR1} that allow one parameter to be adjusted freely. Another one independent parameter of the model is either $F_2(0)$ or $F_3(0)$. Besides, electrodynamic parameters $F_1(0)$, $F_2(0)+F_3(0)$ and the scale $\Lambda$ can be varied, so that not to go beyond experimental errors. Logarithmic renormalization is taken into account within the models with the simplest one-parameter \eqref{Li1} and two-parameter \eqref{Li2} interpolation functions for $n_1=3$, $n_2=1$, $n_3=4$. Thereby, the four-pole models used to fit experimental data comprise 8 and 11 free parameters, respectively, three of which are constrained within experimental uncertainties. In the following we refer to corresponding fits as F1 (one-parameter interpolation functions) and F2 (two-parameter interpolation functions).

The adjusted parameters are set out in Table~\ref{tab:param32}. The corresponding curves are depicted in Figs.~\ref{fig:Delta(1232)AAS}, \ref{fig:Delta(1232)REMRSM}, \ref{fig:Delta(1232)GM}, \ref{fig:Delta(1232)FFF} in comparison with experimental data points collected from the papers \cite{un-06,fr-99,ka-01,st-06,st-98,st-93,PDG,ti-04,az-05a,sp-05,el-06,ke-05}; the curves of the magnetic transition form factor $G_\text{M}^* (Q^2)$ and the ratio $R_{\text{EM}} (Q^2)$ are displayed up to $15 \text{ GeV}^2$, since the high-energy measurements in this region are proposed by JLab Hall~C collaboration~\cite{st-01}.

Distinctions between the models with \eqref{Li1} and \eqref{Li2} are clearly seen in Figs.~\ref{fig:Delta(1232)REMRSM} and \ref{fig:Delta(1232)FFF} displaying the ratios $R_{\text{EM}}$, $R_{\text{SM}}$ and extracted form factors. The model F1 with one-parameter logarithmic renormalization tends to underestimate significantly the magnitude of the Mainz data \cite{st-06,sp-05} on $R_{\text{SM}}$ in the quasistatic domain. While the nine-parameter fit F2 does not suffer from this flaw, it, however, predicts electric quadrupole moment to change sign at $5.5 \text{ GeV}^2$, which contradicts the highest to date JLab data point indicating no sign change up to $6 \text{ GeV}^2$ \cite{un-06}.

It should be noted at this point, that any realization of VMD model involving logarithmic renormalization requires putting forward reliable hypothesis about the way to introduce logarithmic corrections. In the developed framework, such an arbitrary treatment of logarithmic interpolation functions originates in part from the supposition of the values of the exponents $n_1$, $n_2$, $n_3$. This supposition is necessary only until the proper calculations of the helicity amplitude asymptotes including logarithmic corrections are carried out. However, only an improving experimental data seems to be an ultimate solution to the problem that could rule out some interpolation formulas and reduce discrepancy between fits making use of the rest allowed ones. In this regard, the quasistatic domain is as important as proposed high-energy JLab measurements~\cite{st-01}. For example, the current errors of the helicity amplitude extraction do not exclude two types of $Q^2$-evolution of the form factors $F_2$ and $F_3$ near the photon point: monotonous falloff and the $Q^2$-behavior with the derivative changing sign. It may be shown that the model with $K=4$ and logarithmic renormalization \eqref{Li1} is capable of reproducing the first regime only. Therefore, the future measurements could prove such a model to be inadequate in the limit of large distances. Also the experiments at small $Q^2$ might reveal the sign change of the form factors $F_2$ and $F_3$. The zero of the form factors can be reproduced only in the model with the number of $\rho$-mesons $K \geqslant 5$.

The values of the longitudinal amplitude at photon point $S_{1/2} (0)$ are $0.011 \text{ GeV}^{-1/2}$ for the fit F1 and $0.017 \text{ GeV}^{-1/2}$ for the fit F2.

A good agreement of the four-pole models with experimental data ($\chi^2/\text{DOF} \approx 1.6-2.0$) testifies that physics of the transition form factors can be formulated in terms of the QCD-inspired VMD model which deals with all excited states of the $\rho(770)$ and involves logarithmic renormalization and SRs between parameters of meson spectrum. It is remarkable that this good fit is possible in the model with the minimal number of free parameters.

\begin{table}
\caption{\label{tab:param32}Fit parameters (spin-3/2 resonances). Dependent parameters are tabulated in the bottom part of the table ($a^{(\rho,p)}_{3k}$ are not presented, since $a^{(\rho,p)}_{2k} = a^{(\rho,p)}_{3k}$ in the 4-pole model). F1 is a fit with logarithmic functions \eqref{Li1}; F2 is a fit with logarithmic functions \eqref{Li2}.}
\begin{ruledtabular}
\begin{tabular}{ccccc}
& \multicolumn{2}{c}{$\Delta(1232)$} & \multicolumn{2}{c}{$N(1520)$} \\
\cline{2-3}\cline{4-5}
&F1&F2&F1&F2\\
\hline
$\chi^2/$DOF & $\phm 1.98 \phn $ & $\phm 1.63 \phn$ & \begin{tabular}{c} $\phm 3.33$\hphantom{\footnotemark[1]} \\ $\phm 2.00$\footnotemark[1] \end{tabular} & \begin{tabular}{c} $\phm 3.34$\hphantom{\footnotemark[1]}\\ $\phm 1.62$\footnotemark[1] \end{tabular} \\
\hline
$a^{(\rho,p)}_{14}$ & $-0.512 $ & $-1.240$ & $\phm 0.565$ & $-0.270$ \\
$F_1^{(p)}(0)$ & $\phm 2.046$ & $\phm 2.052$ & $\phm 2.970$ & $\phm 2.979$ \\
$F_2^{(p)}(0)$ & $-1.382 $ & $-0.898$ & $-1.044$ & $-1.038$ \\
$F_2^{(p)}(0)+F_3^{(p)}(0)$ & $-1.032 $ & $-0.782$ & $-1.635$ & $-1.635$ \\
$\Lambda$ & $\phm 0.190 $ & $\phm 0.190$ & $\phm 0.190 $ & $\phm 0.238 $ \\
$C^{(\rho,p)}_1$ & $\phm 0.010$ & --- & $\phm 0.014$ & --- \\
$C^{(\rho,p)}_2$ & $\phm 0.021$ & --- & $\phm 0 \hphantom{.000}$ & --- \\
$C^{(\rho,p)}_3$ & $\phm 0.003$ & --- & $\phm 0 \hphantom{.000}$ & --- \\
$h^{(\rho,p)}_1$ & --- & $-0.007$ & --- & $\phm 0.006$ \\
$k^{(\rho,p)}_1$ & --- & $\phm 0.014$ & --- & $\phm 0.016$ \\
$h^{(\rho,p)}_2$ & --- & $-0.338$ & --- & $\phm 0 \hphantom{.000}$ \\
$k^{(\rho,p)}_2$ & --- & $\phm 0.053$ & --- & $\phm 0 \hphantom{.000}$ \\
$h^{(\rho,p)}_3$ & --- & $-0.278$ & --- & $\phm 0 \hphantom{.000}$ \\
$k^{(\rho,p)}_3$ & --- & $\phm 0.052$ & --- & $\phm 0 \hphantom{.000}$ \\
\hline
$a^{(\rho,p)}_{11}$ & $\phm 1.870 $ & $\phm 2.041$ & $\phm 1.564$ & $\phm 1.683$ \\
$a^{(\rho,p)}_{12}$ & $-2.113 $ & $-3.122$ & $-0.260$ & $-0.903$ \\
$a^{(\rho,p)}_{13}$ & $\phm 1.756 $ & $\phm 3.321$ & $-0.870$ & $-0.049$ \\
$a^{(\rho,p)}_{21}$ & $\phm 2.101 $ & $\phm 2.101$ & $\phm 2.142$ & $\phm 2.142$ \\
$a^{(\rho,p)}_{22}$ & $-3.586 $ & $-3.586$ & $-3.410$ & $-3.410$ \\
$a^{(\rho,p)}_{23}$ & $\phm 4.033 $ & $\phm 4.033$ & $\phm 3.146$ & $\phm 3.146$ \\
$a^{(\rho,p)}_{24}$ & $-1.548 $ & $-1.548$ & $-0.879$ & $-0.879$ \\
\end{tabular}
\end{ruledtabular}
\footnotetext[1]{This is the value of $\chi^2/$DOF recalculated with data points at $1 \text{ GeV}^2$ and $1.45 \text{ GeV}^2$ being excluded from the data set. These points disagree significantly with others, which is seen in Fig.~\ref{fig:N(1520)AAS}.}
\end{table}

%==============================================================================================
%==============================================================================================
%==============================================================================================

\subsubsection{Problem of the transition to pQCD}

\begin{figure*}
	\includegraphics[width=0.49\linewidth]{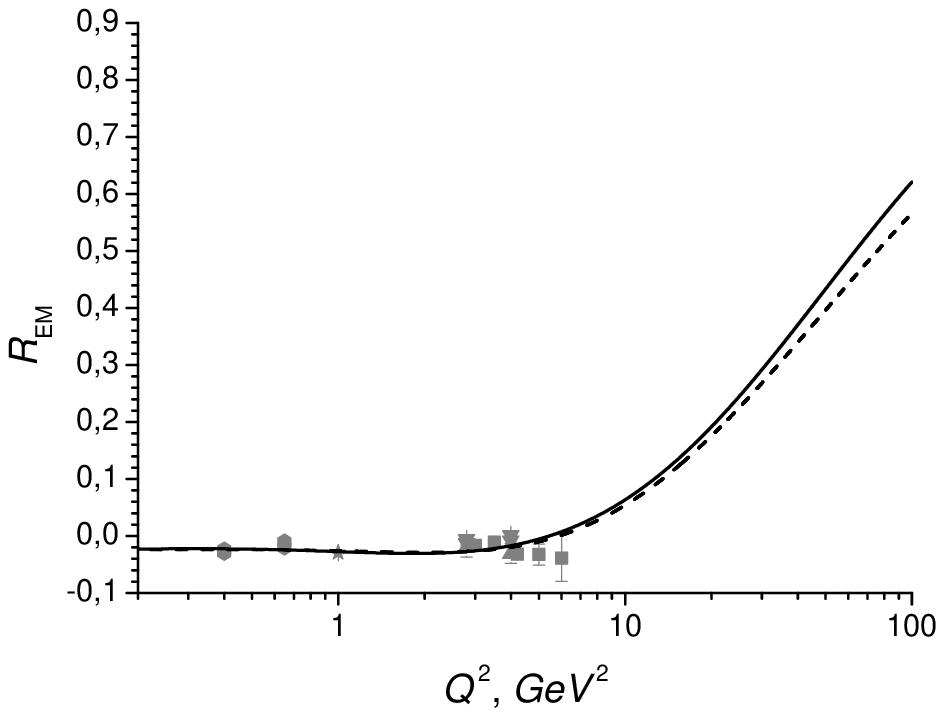}
	\includegraphics[width=0.49\linewidth]{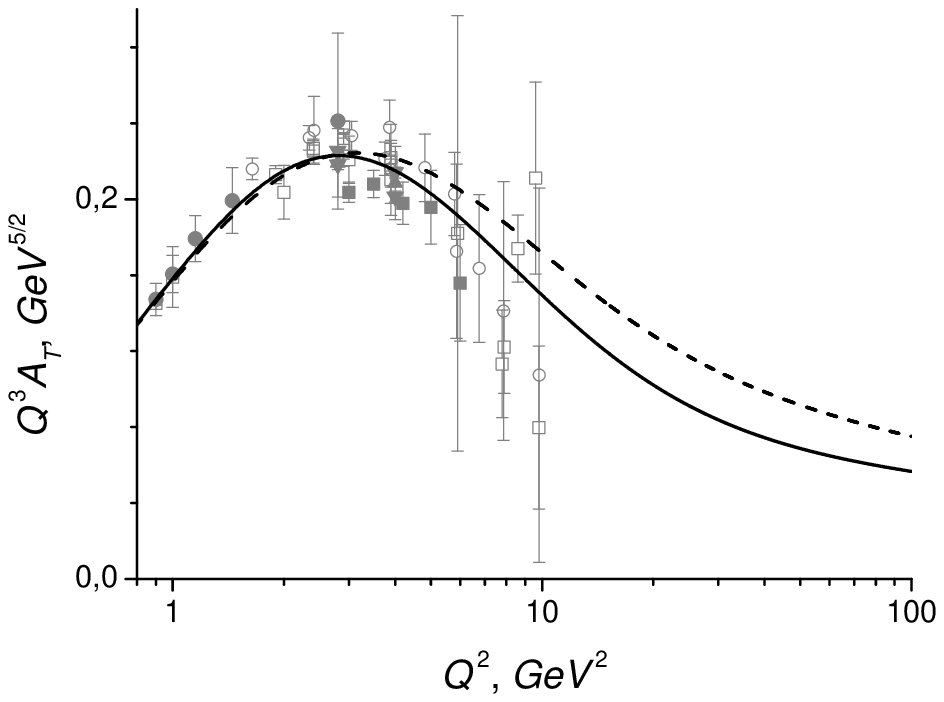}
	\caption{\label{fig:Delta(1232)Asymptotes}Asymptotic behavior of the ratio $R_\text{EM}$ and total transverse amplitude $A_T$ for the $\Delta(1232)$. Curves and data are the same as in Fig.~\ref{fig:Delta(1232)AAS}.}
\end{figure*}

The challenge to observe the onset of asymptotic evolution of resonant helicity amplitudes is one of the goals inspiring experimentalists to carry out high-$Q^2$ measurements~\cite{st-01}. However, the transition to perturbative domain is unlikely to manifest itself anyway in the current world data base on $N\Delta(1232)$-transition form factors. This fact is clearly exemplified by recent exclusive JLab data on $R_{\text{EM}}$ \cite{un-06} and inclusive SLAC data \cite{st-93,st-98} on transverse helicity amplitude $A_T$. JLab results on $R_{\text{EM}}$ depicted in Fig.~\ref{fig:Delta(1232)REMRSM} evidence that this ratio remains small and negative up to $6 \text{ GeV}^2$, while the perturbative asymptote is $R_{\text{EM}} \to +1$. The inclusive data are obtained up to almost $10 \text{ GeV}^2$ and exhibit a trend to decrease more rapidly than $1/Q^3$ predicted by pQCD, though experimental uncertainties are quite considerable. This is readily seen on the right panel of Fig.~\ref{fig:Delta(1232)Asymptotes} that shows transverse amplitude normalized by its asymptote.

The studies of the transition form factors for $Q^2 < 14 \text{ GeV}^2$ proposed by Jefferson Laboratory~\cite{st-01} seem to be of great importance as the meeting ground between predictions made by both baryon-meson and quark-parton physics, which is a new side of quark-hadron duality. It is a well-established fact that asymptotic quark-parton description of inclusive deep-inelastic $eN$-scattering is adequate for $Q^2 \gg T_g^{-2}$, where $T_g =(1.2\,-\,1.5\text{ GeV})^{-1}$ is a space-time scale of nonperturbative quark-gluon fluctuations. However, the transition to pQCD in exclusive resonant process $eN \to e\Delta$ is shifted to higher momentum transfers, which is clearly illustrated by the data depicted in Figs.~\ref{fig:Delta(1232)Asymptotes}. This could be explained by qualitative estimates as follows. Energy transfer from electron to quark (parton) $\Delta E\sim Q^2/2M_N$ should be shared equally between all valence quarks participating in exclusive process and all the quark energy-exchange subprocesses must be hard. Hence, pQCD is the correct description for exclusive reactions in the region of $Q^2 \gg 3 T_g^{-2}\simeq 4.5\,-\,7.5 \text{ GeV}^2$. This does not contradict the current experimental data. Moreover, it is just the domain where VMD model discussed in this paper predicts $R_{\text{EM}}$ to cross zero and rise gradually. It could be regarded as a signal of the transition to pQCD (see Fig.~\ref{fig:Delta(1232)Asymptotes}, left panel).

In baryon-meson physics transition form factors can be described in the framework of the dispersion relation approach or its simplest realization --- the QCD-ispired VMD model. Form factors represented as dispersionlike expansions have correct pQCD asymptotic behavior by the construction, but do not approach them in the region $Q^2<10 \text{ GeV}^2$. Indeed, expanding dispersionlike form factors in inverse powers of $Q^2$ provides a quantitative criterion for the transition to the asymptotic domain:
\begin{align}
	Q^2 &\gg \Lambda_{\text{H} \to \text{Q}}^2
	= \nonumber\\
	&\sup_{\genfrac{}{}{0pt}{1}{n \geqslant p_\alpha+1}{\alpha=1,2,3}} \left[ \sum_{k=1}^{K} a^{(\rho)}_{\alpha k} m^{2 n}_{(\rho)k} \biggl/ \sum_{k=1}^{K} a^{(\rho)}_{\alpha k} m^{2 p_\alpha}_{(\rho)k} \right]^{\genfrac{}{}{}{1}{1}{n-p_\alpha}},
\end{align}
where $p_1=3$, $p_2 = p_3 = 4$ [see Eq.~\eqref{FQCD}]. Thereby, the scale of the transition to pQCD $\Lambda_{\text{H} \to \text{Q}}$ is determined in the VMD model by properties of the interactions between baryons and vector mesons and, certainly, by the structure of the vector-meson spectrum. The quark-hadron duality as an agreement of the predictions by pQCD and baryon-meson models implies the following relation
\begin{equation}
	3 T_g^{-2} \simeq \Lambda_{\text{H} \to \text{Q}}^2.
\end{equation}

The discussed models involving four vector mesons and logarithmic renormalization of the type \eqref{Li1} and \eqref{Li2} provide $\Lambda_{\text{H} \to \text{Q}}^2=9.24 \text{ GeV}^2$ and $\Lambda_{\text{H} \to \text{Q}}^2=9.67 \text{ GeV}^2$ respectively. However, one is forced to accept the fact that these values are affected by the aforementioned intrinsic drawbacks of the model and large experimental errors, especially, at high momentum transfers. But we believe that theoretical refinement of the model and, what is likely to be even more important, future high\nobreakdash-\hspace{0pt}$Q^2$ exclusive experiments followed by extraction of all helicity amplitudes would improve the situation substantially. In this regard it is worth mentioning that the four-pole models are able to reproduce quite different rates of the transition to pQCD, which can be proved by adding some hypothetic experimental data to the current data base and fitting to it. This fact makes us expect that future measurements of transition form factors will neither constrain the range of validity of the QCD-inspired VMD model nor reduce the overall quality of fit.

%==============================================================================================
%==============================================================================================
%==============================================================================================

\subsection{The second resonance region}\label{The second resonance region}

The second resonance region covers the $W$ range between approximately 1.4 GeV and 1.6 GeV. It includes three isospin $1/2$ states $N(1440)$, $N(1520)$, $N(1535)$. Though both $\rho$- and $\omega$-mesons contribute to the excitation of these baryons, currently there is no measurements of neutron helicity amplitudes, except for the photoproduction data~\cite{PDG}. Thus, in the framework developed, it is hardly possible to distinguish reliably isovector and isoscalar contributions to the form factors. Because of this reason, in the following we neglect singlet-triplet mass splitting and suppose that $\rho$- and $\omega$-mesons propagate in the nucleon medium identically, i.e., $L^{(\rho)}_\alpha \equiv L^{(\omega)}_\alpha$. In such a model, proton transition form factors depend on half as many independent parameters as the form factors \eqref{Fvdm1} and \eqref{Gvdm1}:
\begin{equation}\label{Fvdm1vs}
	N(1520):\quad
	F^{(p)}_\alpha (Q^2) =
	\frac{F^{(p)}_\alpha (0)}{L_\alpha^{(p)}(Q^2)} \sum_{k=1}^K \frac{a_{\alpha k}^{(p)} m_k^2}{Q^2+m_k^2};
\end{equation}
\begin{align}\label{Gvdm1vs}
	N(1440)&,\; N(1535):\nonumber
	\\
	&\qquad G^{(p)}_\alpha (Q^2) =
	\frac{G^{(p)}_\alpha (0)}{L_\alpha^{(p)}(Q^2)} \sum_{k=1}^K \frac{b_{\alpha k}^{(p)} m_k^2}{Q^2+m_k^2},
\end{align}
where $L^{(p)}_\alpha \equiv L^{(\rho)}_\alpha \equiv L^{(\omega)}_\alpha$ is a logarithmic interpolation function of type \eqref{Li1} or \eqref{Li2}, $m_k^2 = (m_{(\omega)k}^2 + m_{(\rho)k}^2)/2$ are vector meson masses averaged for each singlet-triplet family from Table~\ref{tab:mesons}, and the expansion parameters $a^{(p)}_k$ satisfy the same set of the SRs (\ref{SR1})--(\ref{SR4}) as $a^{(\rho)}_k$ and $a^{(\omega)}_k$ do.

In spite of the simplifications described above, the form factors \eqref{Fvdm1vs} and \eqref{Gvdm1vs} are found to provide a good fit of the existing data in the second resonance region.

%==============================================================================================
%==============================================================================================
%==============================================================================================

\subsubsection{The $N(1520)$}

Since the resonance $N(1520)$ possesses spin $3/2$, we use the four-pole model \eqref{F1F2F3}. Logarithmic renormalization is taken into account by means of one-parameter \eqref{Li1} and two-parameter \eqref{Li2} interpolation functions (corresponding fits are denoted F1 and F2, respectively) for $n_1=3$, $n_2=1$, $n_3=4$. In both cases, logarithmic dependency of the spin-flip form-factors $F_2^p$, $F_3^p$ could be neglected as it does not manifest itself in the fit to the data in the experimentally acceptable domain $Q^2 < 4 \text{ GeV}^2$. Therefore, the models used to fit the experimental data comprise 6 and 7 adjustable parameters: two electrodynamic parameters $F_1^p(0)$, $F_2^p(0)+F_3^p(0)$ constrained by photoproduction data and one parameter $F_2^p(0)$ varied freely; an expansion coefficient $a^{(p)}_{14}$; QCD-scale $\Lambda = 0.19-0.24 \text{ GeV}$; and one or two parameters of the interpolation function~$L_1^{(p)}$.

The fit results are presented in Table~\ref{tab:param32} and Figs.~\ref{fig:N(1520)AAS}, \ref{fig:N(1520)FFF}. The large values of $\chi^2$/DOF is likely to be attributed to the discrepant data points~\cite{ti-04} in the region between $Q^2 = 1 \text{ GeV}^2$ and $Q^2 = 2 \text{ GeV}^2$.

It should be noted that the good fits to the experimental data on helicity amplitudes of two spin-vector states $\Delta(1232)$ and $N(1520)$ are obtained in the unified approach, based on four-pole dispersionlike form-factor expansions \eqref{F1F2F3} satisfying SRs (\ref{SR1})--(\ref{SR4}). This is an evidence for validity of the VMD model in physics of high-spin nucleon excitations.

%==============================================================================================
%==============================================================================================
%==============================================================================================

\subsubsection{The $N(1535)$}\label{The N(1535)}

The resonances $N(1535)$ and $N(1440)$ are spin-$1/2$ nucleon excitations. We fit the data by means of four- and five-pole models (the corresponding fits are denoted F1 and F2, respectively) with two-parameter renormalization \eqref{Li2}, because the interpolation of the type \eqref{Li1} is found to be unsatisfactory to reproduce $Q^2$-dependence of the form factors. So, the fit models F1 and F2 introduce 8 and 10 free parameters: two electrodynamic parameters $G_1^p(0)$ and $G_2^p(0)$ (the second one is constrained by photoproduction data), five parameters of functions $L_1^{(p)}$ and $L_2^{(p)}$ (QCD scale $\Lambda$ is varied between $0.19 \text{ GeV}$ and $0.24 \text{ GeV}$), one (in the case of four-meson fit F1) or three (in the case of five-meson fit F2) of the expansion coefficients $b_{\alpha k}^{(p)}$.

The available data points \cite{ti-04,az-05a,la-06,st-93,th-01,ar-99,PDG} are divided into two samples fitted separately. The first data sample S1 is all the data with the analyses \cite{az-05a,la-06} by I.~G.~Aznauryan \textit{et~al.} being excluded. The second one denoted S2 includes the PDG average at photon point \cite{PDG}, inclusive data from \cite{st-93} in the region $Q^2 > 5 \text{ GeV}^2$, where exclusive experiments have not yet been carried out, and the most recent exclusive JLab data \cite{ar-99,az-05a,la-06}. The reason to fit these data samples apart is that analysis fullfilled by L.~Tiator \textit{et al.} \cite{ti-04} predicts slightly more rapid falloff of transverse amplitude and more substantial rise of longitudinal one than the data points from \cite{az-05a,la-06} exhibit (Fig.~\ref{fig:N(1535)AS}). In the model based on vertex~\eqref{G12}, this contradiction manifests itself in the most obvious way as the discrepancy in the extracted data on the form factor $G_1^p$, depicted in Fig.~\ref{fig:N(1535)GG}. Nevertheless, the four- and five-pole models provide good fits to both data samples.

The fit parameters are set out in Table~\ref{tab:param12}. The corresponding curves are shown in Figs.~\ref{fig:N(1535)AS} and \ref{fig:N(1535)GG}. It is of some interest that the fits to both data samples give the value of $b_{15}^{(p)}$ to be an order of magnitude less than $b_{14}^{(p)}$. It could be regarded as evidence that only the first four meson families contribute to the form factor $G_1^p$ of the resonance $N(1535)$.

%==============================================================================================
%==============================================================================================
%==============================================================================================

\subsubsection{The $N(1440)$}\label{The N(1440)}

To fit the experimental data on the Roper resonance $N(1440)$, we make use of the same model as that described in the previous subsection~\ref{The N(1535)}. However, unlike the form factors of the transition $N \to N(1535)$, both nucleon-to-Roper form factors cross zero in the region $Q^2 < 0.7 \text{ GeV}^2$ (Fig.~\ref{fig:N(1440)GG}). To incorporate this effect is only possible in the models involving at least five mesons, since the simplest four-pole form factors with correct power pQCD-asymptotes \eqref{G1ass} are predicted by the model with meson masses from~\cite{PDG} to be monotonous and to conserve the sign.

As in the case of the $N(1535)$, there is some discrepancy between results of the helicity-amplitude extraction made in the framework of the MAID model \cite{ti-04} and JLab UIM \cite{az-05a, az-05b}. For instance, while the analysis \cite{az-05a, az-05b} indicates that the form factor $G_2^p$ crosses zero between $0.4 \text{ GeV}^2$ and $0.65 \text{ GeV}^2$, MAID calculations \cite{ti-04} shift the sign change to the domain between $0.75 \text{ GeV}^2$ and $0.9 \text{ GeV}^2$. That's why we divide the data points into two samples and fit to them separately. Both samples include PDG averages at photon point \cite{PDG} and the data from \cite{la-06} for $Q^2>1.5 \text{ GeV}^2$, but in the region $Q^2<1.5 \text{ GeV}^2$ the first sample S1 takes into account just the analysis \cite{ti-04}, and the second one S2  includes the data from~\cite{az-05a, az-05b}.

The adjusted parameters are tabulated in Table~\ref{tab:param12}. The corresponding helicity amplitudes and extracted form factors are depicted in Figs.~\ref{fig:N(1440)AS} and \ref{fig:N(1440)GG}.

\begin{table}
\caption{\label{tab:param12}Fit parameters (spin-1/2 resonances). Dependent parameters are tabulated in the bottom part of the table. S1, S2 are data samples introduced in the text. F1 is a four-pole fit; F2 is a five-pole fit.}
\begin{ruledtabular}
\begin{tabular}{ccccccc}
& \multicolumn{4}{c}{$N(1535)$} & \multicolumn{2}{c}{$N(1440)$} \\
\cline{2-5}\cline{6-7}
&\multicolumn{2}{c}{S1}&\multicolumn{2}{c}{S2}&S1&S2\\
\cline{2-3}\cline{4-5}\cline{6-6}\cline{7-7}
&F1&F2&F1&F2&F2&F2\\
\hline
$\chi^2/$DOF & $\phm 2.61 \phn $ & $\phm 2.45 \phn$ & $\phm 0.46 \phn $ & $\phm 0.48 \phn $ & \begin{tabular}{c} $\phm 7.91$\hphantom{\footnotemark[1]} \\ $\phm 3.87$\footnotemark[1] \end{tabular} & $\phm 1.69 \phn$ \\
\hline
$b^{(p)}_{14}$ & $\phm 0.860$ & $\phm 1.082$ & $\phm 1.964$ & $\phm 2.015$ & $-2.215$ & $-2.591$ \\
$b^{(p)}_{15}$ & --- & $\phm 0.048$ & --- & $-0.060$ & $-0.322$ & $-0.632$ \\
$b^{(p)}_{25}$ & --- & $\phm 1.064$ & --- & $\phm 0.872$ & $\phm 5.814$ & $\phm 9.402$ \\
$G_1^{(p)}(0)$ & $\phm 0.417$ & $\phm 0.407$ & $\phm 0.179$ & $\phm 0.181$ & $-0.052$ & $-0.034$ \\
$G_2^{(p)}(0)$ & $\phm 0.210$ & $\phm 0.210$ & $\phm 0.210$ & $\phm 0.214$ & $-0.241$ & $-0.252$ \\
$\Lambda$ & $\phm 0.240 $ & $\phm 0.240$ & $\phm 0.240$ & $\phm 0.240$ & $\phm 0.240 $ & $\phm 0.240 $ \\
$h^{(p)}_1$ & $-0.001$ & $\phm 0.006$ & $-0.309$ & $-0.304$ & $-0.579$ & $-0.660$ \\
$k^{(p)}_1$ & $\phm 0.009$ & $\phm 0.012$ & $\phm 0.057$ & $\phm 0.054$ & $\phm 0.095$ & $\phm 0.121$ \\
$h^{(p)}_2$ & $-0.491$ & $-0.528$ & $-0.492$ & $-0.515$ & $-0.555$ & $-0.479$ \\
$k^{(p)}_2$ & $\phm 0.062$ & $\phm 0.070$ & $\phm 0.065$ & $\phm 0.068$ & $\phm 0.082$ & $\phm 0.087$ \\
\hline
$b^{(p)}_{11}$ & $\phm 1.446$ & $\phm 1.294$ & $\phm 1.005$ & $\phm 1.064$ & $\phm 3.105$ & $\phm 3.667$ \\
$b^{(p)}_{12}$ & $\phm 0.384 $ & $\phm 1.156$ & $\phm 2.793$ & $\phm 2.545$ & $-8.254$ & $-10.93$ \\
$b^{(p)}_{13}$ & $-1.691$ & $-2.580$ & $-4.762$ & $-4.564$ & $\phm 8.687$ & $\phm 11.49$ \\
$b^{(p)}_{21}$ & $\phm 2.142$ & $\phm 2.563$ & $\phm 2.142$ & $\phm 2.487$ & $\phm 4.441$ & $\phm 5.860$ \\
$b^{(p)}_{22}$ & $-3.410$ & $-7.034$ & $-3.410$ & $-6.380$ & $-23.21$ & $-35.43$ \\
$b^{(p)}_{23}$ & $\phm 3.146$ & $\phm 9.868$ & $\phm 3.146$ & $\phm 8.655$ & $\phm 39.87$ & $\phm 62.54$ \\
$b^{(p)}_{24}$ & $-0.879$ & $-5.461$ & $-0.879$ & $-4.634$ & $-25.92$ & $-41.37$ \\
\end{tabular}
\end{ruledtabular}
\footnotetext[1]{This is the value of $\chi^2/$DOF recalculated with data points at $0.525 \text{ GeV}^2$ and $1.45 \text{ GeV}^2$ being excluded from the data sample. These points disagree significantly with others from the sample S1, which is seen in Fig.~\ref{fig:N(1440)AS}.}
\end{table}

%==============================================================================================
%==============================================================================================
%==============================================================================================

\section{Conclusion}\label{Conclusion}

We have investigated $NR${\nobreakdash-\hspace{0pt}}form factors in the first and second resonance regions, utilizing effective-field theory with 4--5 explicit vector-meson degrees of freedom. Transition form factors in the model comprise 6--10 free parameters for each resonance which have been fitted to experimental data. All these parameters have clear physical meaning (low-energy electromagnetic constants, meson-baryon couplings, and phenomenological parameters of logarithmic renormalization).

This QCD-inspired VMD model is in good agreement with the data available on resonant helicity amplitude in the first and second resonance regions. This success makes us believe that the model, though being phenomenological, provides an insight into the $Q^2$-evolution of nucleon-to-resonance transitions. The basic physical ideas of the approach are as follows:

\begin{enumerate}
\item The photon, propagating in the inside-nucleon medium, excites all the modes of a hadronic string, carrying photon quantum numbers $J^{PC}=1^{--}$. Thus, all the vector mesons should be, in principle, incorporated into the VMD model, which makes the form-factors be dispersionlike expansions with poles at meson masses.
\item Short-distance quark-gluon processes contribute to the hadronization of a photon into intermediary mesons inside nucleon, i.e., at $Q^2 > R_N^{-2} = (0.2 \text{ GeV} )^2$. The VMD model takes into account the small-scale dynamics by effective logarithmic renormalization of electrodynamic coupling constants.
\item The VMD model should be reconciled with pQCD, commonly believed to be the \textit{ab initio} treatment of resonant electroproduction at high momentum transfers. To attain requisite asymptotic behavior of the form-factors is possible by imposing linear superconvergence relations on the parameters of the vector-meson spectrum. Besides, logarithmic renormalization of meson-baryon parameters is essential at this point, as it allows to include logarithmic corrections to pQCD-asymptotes and to make the asymptotes of spin-flip and non-spin-flip phenomenological form factors match with their counterparts at quark level.
\end{enumerate}

It should be noted that inclusion of all the vector mesons appears to be impossible in the framework of the VMD model tested in this paper, as it could overparametrize fit to the data. However, even the simplest models with four and five lightest vector mesons, which introduce only one and three independent meson-baryon coupling, respectively, is found to be in accord with all the experimental data analyzed. In our opinion, it supports the notion that the aforementioned effects make a major contribution to $Q^2$-evolution of the nonstrange resonance excitation.

Further work, however, needs to be done to improve vector-meson-dominance model of the nucleon transition form factors. This improvement should include both theoretical refinement and new experiments. From the theoretical point of view, it seems to be important to test alternative Lagrangians of nucleon interactions with high-spin resonances \cite{futurepub} and to carry out pQCD-calculations of logarithmic corrections to form-factor asymptotes. Also, it seems reasonable to extend the model to be directly compared with experimental data on $eN$-scattering observables, for the extraction of helicity amplitudes is known to be model dependent \cite{da-91}. Future experiments in both quasistatic and high-$Q^2$ regions will provide important information imposing constraints on phenomenological logarithmic renormalization and, especially, on the contributions of nonleading pQCD-logarithms.

%==============================================================================================
%==============================================================================================
%==============================================================================================

\begin{acknowledgments}
We are grateful to A.~V.~Beylin and V.~I.~Kuksa for their reading the draft of this paper, useful comments, and invaluable criticisms.
\end{acknowledgments}

%==============================================================================================
%==============================================================================================
%==============================================================================================

\appendix*

%==============================================================================================
%==============================================================================================
%==============================================================================================

\section{Helicity-amplitude phases}

Helicity amplitudes are defined as matrix elements of the electromagnetic current operator calculated between the initial nucleon and final resonance states. As the polarized states of spinor fields always contain arbitrary phase, helicity amplitudes are also defined up to phases, that cannot be calculated in the extraction based on the simple cross section formula \eqref{sig}. That's why we put forward three additional empirical criteria that could fix amplitude phases:
\begin{enumerate}
\item electrodynamic parameters should be within a factor of ten;
\item form-factor expansion coefficients should have an order of magnitude of 1 or less;
\item convergence and quality of the fit.
\end{enumerate}

These criteria are, obviously, ``fit-dependent", i.e., phases might depend upon specific features of the model, such as the choice of logarithmic interpolation functions, the meson spectrum, etc. However, we found them to be able to fix the phases reliably, at least for the resonances $\Delta(1232)$, $N(1520)$, $N(1535)$.

In the scope of this paper, only the ratios of the phases are important and one of the amplitude phases can be fixed arbitrarily. We choose the phase of the amplitude $A_{1/2}$.

%==============================================================================================
%==============================================================================================
%==============================================================================================

\subsection{The $\Delta(1232)$}

In the case of the $\Delta(1232)$, the ratio of the phases of the transverse amplitudes is determined by the first criterion. Indeed, amplitudes \eqref{A32} and \eqref{A12} give $F_1^{(p)}(0) \approx 2.068$, $F_2^{(p)}(0)+F_3^{(p)}(0) \approx 1.185$. If someone changes the sign of the amplitude \eqref{A12}, the values of electrodynamic parameters change to $F_1^{(p)}(0) \approx 0.069$, $F_2^{(p)}(0)+F_3^{(p)}(0) \approx 11.62$. This option does not fulfill the first criterion. Besides, it is found impossible to fit the experimental data in the four-pole model with such phases, while this model can reproduce observed $Q^2$-evolution of helicity amplitudes \eqref{A32}--\eqref{S12}.

The sign of the longitudinal amplitude $S_{1/2}$ is fixed by the third criterion, since the four-pole model with the sign of the $S_{1/2}$ opposite to that of \eqref{S12} crucially underestimates the ratio $R_{\text{SM}}$ predicting it to be of an order of 0.001 or less for $Q^2 > 3 \text{ GeV}^2$.

%==============================================================================================
%==============================================================================================
%==============================================================================================

\subsection{The $N(1520)$}

For $N(1520)$, the first criterion is not effective, since all the isovector and isoscalar electromagnetic parameters are of the same order of magnitude, regardless of the ratio of the transverse amplitudes phases. Therefore, phases of the amplitudes \eqref{A32}--\eqref{S12} are chosen by the quality of the fits. All the alternatives to formulas \eqref{A32}--\eqref{S12} lead to a several-fold increase in $\chi^2$/DOF compared to the best values, obtained in the model \eqref{A32}--\eqref{S12}.

%==============================================================================================
%==============================================================================================
%==============================================================================================

\subsection{The $N(1535)$}

The values of $\chi^2$/DOF obtained in the fits with the sign of $S_{1/2}$ being opposite to that of \eqref{S1212} are 1.2 (S1-F1), 1.34 (S1-F2), 2.67 (S2-F1), 2.38 (S2-F2). The first two values are more than twice as large as those set out in Table~\ref{tab:param12}. However, the difference in the quality of the fits to the data sample S2 is subtle. Nevertheless, these fits converge for some of the form-factor expansion coefficients being of an order of 10, which dissatisfies the second criterion.

%==============================================================================================
%==============================================================================================
%==============================================================================================

\subsection{The $N(1440)$}

The case of the Roper resonance is more controversial. The data on helicity amplitudes is quite sparse and discrepant, which results in ambiguity of the amplitude signs. The Eqs.~\eqref{A1212} and \eqref{S1212} provide a much better value of $\chi^2$/DOF in the fit to the data sample S2 than the model with the opposite sign of $S_{1/2}$ does. But, the contrary is found in the fits to the data sample S1 with the points at $0.525 \text{ GeV}^2$ and $1.45 \text{ GeV}^2$ being excluded. So, the choice of amplitude signs for the $N(1440)$ essentially depends on the particular sample of experimental data. We choose the signs which allow the best fit to the sample S2, since it includes the most recent results of helicity amplitude extraction \cite{az-05a,az-05b,la-06} from the data on one and two pion electroproduction.

%==============================================================================================
%==============================================================================================
%==============================================================================================

\bibliography{vervolch}

\end{document}